\documentclass[%
 reprint,
 superscriptaddress,
 preprintnumbers,
 nofootinbib,
 amsmath,amssymb,
 aps
]{revtex4}

\usepackage[T1]{fontenc} 
\usepackage[utf8]{inputenc}
\usepackage[USenglish]{babel}
\usepackage{soul}
\usepackage{cancel}
\usepackage[normalem]{ulem}
\usepackage{polski}
\usepackage{dsfont}

\usepackage{multirow}
\usepackage[dvipsnames]{xcolor}
\usepackage{graphicx} 
\usepackage{mathtools}
\usepackage{float}
\usepackage{gensymb}
\usepackage{booktabs}

\usepackage{xcolor} 
\usepackage{amsmath,amsfonts,amssymb} 
\usepackage{siunitx}
\usepackage{tensor}
\usepackage{caption} 
\usepackage{subcaption}
\usepackage{dsfont}
\usepackage{braket}
\usepackage{dcolumn}
\usepackage{bm}
\DeclareMathAlphabet{\mathpzc}{OT1}{pzc}{m}{it}
\usepackage{url}
\usepackage{slashed}

\usepackage[bookmarks=true, pdfpagelabels=true, linkcolor=black, urlcolor=black, citecolor=blue, anchorcolor=blue]{hyperref}

\usepackage[capitalise,english]{cleveref}
\newcommand{\be}{\begin{equation}}
\newcommand{\ee}{\end{equation}}
\newcommand{\beq}{\begin{equation}}
\newcommand{\eeq}{\end{equation}}

\def\be{\begin{equation}}
\def\ee{\end{equation}}
\def\ba{\begin{eqnarray}}
\def\ea{\end{eqnarray}}

\makeatletter
\@ifundefined{textcolor}{}
{
 \definecolor{BLACK}{gray}{0}
 \definecolor{WHITE}{gray}{1}
 \definecolor{RED}{rgb}{1,0,0}
 \definecolor{GREEN}{rgb}{0,1,0}
 \definecolor{BLUE}{rgb}{0,0,1}
 \definecolor{CYAN}{cmyk}{1,0,0,0}
 \definecolor{MAGENTA}{cmyk}{0,1,0,0}
 \definecolor{YELLOW}{cmyk}{0,0,1,0}
}

\makeatother

\begin{document}

\title{
Enhancing anomaly detection with topology-aware autoencoders 
}

\author{Vishal S.\ Ngairangbam}
\email{vishal.s.ngairangbam@durham.ac.uk}
\affiliation{%
Institute for Particle Physics Phenomenology, Department of Physics, Durham University, Durham DH1 3LE, U.K.%
}

\author{Błażej Rozwoda}
\email{b.rozwoda@student.uw.edu.pl}
\affiliation{%
Institute of Theoretical Physics, Faculty of Physics,
University of Warsaw, ul.\ Pasteura 5, PL-02-093 Warsaw, Poland%
}

\author{Kazuki Sakurai}
\email{kazuki.sakurai@fuw.edu.pl}
\affiliation{%
Institute of Theoretical Physics, Faculty of Physics,
University of Warsaw, ul.\ Pasteura 5, PL-02-093 Warsaw, Poland%
}

\author{Michael Spannowsky}
\email{michael.spannowsky@durham.ac.uk}
\affiliation{%
Institute for Particle Physics Phenomenology, Department of Physics, Durham University, Durham DH1 3LE, U.K.%
}

\preprint{IPPP/25/08}

\begin{abstract}

Anomaly detection in high-energy physics is essential for identifying new physics beyond the Standard Model. Autoencoders provide a signal-agnostic approach but are limited by the topology of their latent space. This work explores topology-aware autoencoders, embedding phase-space distributions onto compact manifolds that reflect energy-momentum conservation. We construct autoencoders with spherical ($S^n$), product ($S^2 \otimes S^2$), and projective ($\mathds{RP}^2$) latent spaces and compare their anomaly detection performance against conventional Euclidean embeddings.
Our results show that autoencoders with topological priors significantly improve anomaly separation by preserving the global structure of the data manifold and reducing spurious reconstruction errors. Applying our approach to simulated hadronic top-quark decays, we show that latent spaces with appropriate topological constraints enhance sensitivity and robustness in detecting anomalous events. This study establishes topology-aware autoencoders as a powerful tool for unsupervised searches for new physics in particle-collision data.

\end{abstract}

%\date{\today}

\maketitle
%\begin{bibunit}

\section{Introduction} 

Anomaly detection in high-energy particle physics~\cite{Belis:2023mqs,Kasieczka:2021xcg,Aarrestad:2021oeb,DAgnolo:2018cun,Collins:2018epr,Collins:2019jip,Hajer:2018kqm,DeSimone:2018efk,Andreassen:2020nkr,Nachman:2020lpy,Knapp:2020dde,ATLAS:2020iwa,Dillon:2020quc,CrispimRomao:2020ucc,Cheng:2020dal,Khosa:2020qrz,Mikuni:2020qds,Park:2020pak,Blance:2020ktp,Dorigo:2021iyy,Caron:2021wmq,Hallin:2021wme,Mikuni:2021nwn,dAgnolo:2021aun,Park:2022zov,Hallin:2022eoq,Kasieczka:2022naq,Hao:2022zns,Golling:2023juz,ATLAS:2023azi,Metodiev:2023izu,Sengupta:2023vtm,Cheng:2024yig,Grosso:2024nho,Duffy:2024zog,Das:2024fwo,Craig:2024rlv,Araz:2024lsl,Das:2024eie,Hammad:2024dsn} plays a crucial role in uncovering physics beyond the Standard Model (BSM). Traditional search strategies often rely on predefined hypotheses about new physics signals, limiting their applicability when the signal is unknown. Recent advances in machine learning, particularly unsupervised learning techniques, offer new avenues for anomaly detection in collider data. Autoencoders, which learn to compress and reconstruct data, have emerged as a promising tool~\cite{Farina:2018fyg,Heimel:2018mkt,Roy:2019jae,Cerri:2018anq,Blance:2019ibf,vanBeekveld:2020txa,Batson:2021agz,Dillon:2021nxw,Finke:2021sdf,Atkinson:2021nlt,Govorkova:2021utb,Fraser:2021lxm,Tsan:2021brw,Jawahar:2021vyu,Canelli:2021aps,Ngairangbam:2021yma,Alvi:2022fkk,Atkinson:2022uzb,Dillon:2022mkq,Bhattacherjee:2023evs} due to their ability to detect deviations from learned background distributions. However, their effectiveness is constrained by the topology of the latent space in which the data is embedded \cite{Batson:2021agz}.

In this work, we explore the role of topological priors in enhancing anomaly detection performance using autoencoders. Specifically, we recognize that phase space distributions of final-state particles in high-energy collisions reside on non-trivial manifolds dictated by energy-momentum conservation. When the topology of the background manifold is known \textit{a priori}, constructing an autoencoder with a latent space that shares the same topology ensures a more faithful representation of the data. Such a design minimizes distortions in data reconstruction and enhances the network's ability to distinguish signal from background.

To achieve this, we construct and compare several autoencoders with compact and topologically non-trivial latent spaces. We examine different latent space manifolds and evaluate their impact on anomaly detection in particle collider data. By mapping phase space distributions onto topologically structured latent spaces, we aim to clear topological obstructions that degrade the performance of conventional autoencoders with trivial latent space topology.

Our approach involves constructing a variety of autoencoder architectures with different latent space constraints. We investigate autoencoders with spherical ($S^n$), product ($S^2 \otimes S^2$), and projective ($\mathds{RP}^2$) latent manifolds and compare their reconstruction fidelity to standard Euclidean latent representations. Through extensive testing on both toy datasets and realistic collider simulations, we quantify how different topological priors impact reconstruction error and anomaly detection sensitivity. Our experiments demonstrate that autoencoders with appropriately chosen latent topologies achieve superior separation between background and anomalous events, mainly when the anomaly corresponds to a different intrinsic manifold structure.

We validate our method by applying it to a realistic collider physics scenario using simulated hadronic top-quark decays. We show that leveraging a latent space that matches the intrinsic structure of the background data leads to more reliable anomaly detection, reducing false positive rates and increasing robustness against spurious high reconstruction errors due to topological mismatches. Furthermore, we provide a systematic analysis of the latent space embeddings, demonstrating that topological autoencoders effectively preserve the global structure of the data manifold while enhancing the contrast between background and signal events.

The structure of the paper is as follows: In Section \ref{sec:tprior}, we provide a detailed motivation for incorporating topological priors in anomaly detection. We discuss the mathematical foundation of phase space manifolds in high-energy collisions and their implications for machine learning models. In Section \ref{sec:top_embed}, we present different methods for constructing topologically non-trivial latent spaces, highlighting approaches such as the use of spherical and projective manifolds. Section \ref{sec:toy_clear} demonstrates the impact of these methods on toy datasets, showcasing how different topologies influence reconstruction quality. In Section \ref{sec:latentano}, we apply our approach to a realistic collider physics scenario, using simulated hadronic top-quark decays as a test case. We compare the performance of various autoencoders in detecting anomalous three-body decays. Finally, Section \ref{sec:conclusions} summarises our findings and discusses future directions, including possible extensions to more complex final-state topologies.

%===================================================================
\section{Necessity of Topological Priors}
\label{sec:tprior}
%===================================================================
In this section, we motivate the need for topological priors in the latent space of an autoencoder for unsupervised anomaly detection at particle colliders due to the non-trivial topology of the momentum data manifold.
Intuitively, a $k$-dimensional manifold $\mathcal{M}^k$ is a space which locally resembles a Euclidean space $\mathbb{R}^k$ while it can have non-trivial additional features globally. This resemblance is given a precise meaning via a collection of invertible maps called \emph{charts} defined in any given local neighbourhood of $\mathcal{M}^k$ that satisfy additional conditions to designate the mathematical property under study.  
In physics, the study of manifolds generally presumes \emph{smoothness}, where, in addition to invertibility, one requires each map and its inverse to be infinitely differentiable. Due to the universal approximation property of dense Artificial Neural Networks, where they can approximate continuous functions,\footnote{The proof of universal approximation theorems generally consider the function's domains to be the $N$-dimensional closed unit hypercube. Strictly speaking, this is not homeomorphic to the trivial topology on $\mathbb{R}^N$. However, considering the discrete and finite nature of data, one is always restricted to a compact subset in $\mathbb{R}^N$, which we are assuming is the intrinsic data manifold.} we will relax the condition of smoothness to that of continuity. Therefore, all manifolds considered in our work are \emph{topological manifolds}, where local charts (and their inverse) must be at most continuous. Our primary concern will therefore be to study topological equivalence or \emph{homeomorphisms} where one space can be continuously deformed to the other and vice-versa, specifically, the topology of intrinsic $k$-dimensional phase space manifolds which are not homeomorphic to $\mathbb{R}^k$ and hence possess non-trivial topological features.

Putting aside topological considerations, it is important to understand the relationship between the signal and the background manifold. For instance, when the signal manifold is a submanifold of the background, one expects an autoencoder to have an efficient reconstruction of the signal manifold due to the excellent interpolation capabilities of artificial neural networks.  
Therefore, we will primarily consider the signal manifold to be ``out-of-distribution'' and not fully contained in the background manifold.

\subsection{Topological Obstructions}
Various autoencoders have been studied with varying degrees of effectiveness and use cases for anomaly detection. 
The method assumes the background data that is characterised by $N$ features, $x \in {\mathbb R}^N$, is confined in a $k$-dimensional intrinsic data manifold, ${\cal D}^k$, with $k < N$.
In this case, the data may be represented in an $M$-dimensional latent space, $z \in {\mathbb R}^{M}$, with $k \leq M < N$.
By optimising a large number of tunable parameters $\theta$ in the encoder function $z = f_\theta(x)$ and $\phi$ in the decoder function $x' = g_\phi (z)$, it may be possible to make the Mean Square Error (MSE) loss function arbitrarily small (up to a noise) for all background data points, effectively establishing the input-output identity map $x \to x' = (g_\phi \circ f_\theta)(x) = x$.
For signal data points that live outside the background manifold, reconstruction errors tend to be large, enabling us to detect them as anomalies.  Anomaly detection is less effective for larger latent dimensions since it is easier to represent outliers with many degrees of freedom.   On the other hand, for $M < k$, the latent space cannot describe the background data. One may expect the best anomaly detection to be realised for $M = k$.

It has been shown in Ref.~\cite{Batson:2021agz}, however, that this naive expectation is incorrect when background manifolds possess non-trivial topologies. Specifically, maintaining a trivial topology in a $k$-dimensional latent space can lead to regions in the data manifold $\mathcal{D}^k$ exhibiting high reconstruction error.  
These regions may then be spuriously identified as anomalous samples.  
This issue arises for any non-trivial data manifold that requires more than one chart to faithfully represent its local coordinates in $\mathbb{R}^k$, as a single map cannot globally describe the manifold. 

Mathematically, the $k$-dimensional manifold $\mathcal{D}^k$ is \emph{embedded} in $\mathbb{R}^N$, i.e. a subset of $\mathbb{R}^N$ is topologically identical to $\mathcal{D}^k$ and hence the former is a \emph{submanifold} of $\mathbb{R}^N$.
Let $f_\theta:\mathcal{D}^k\to\mathcal{M}^{k_l}$ denote the encoder map,\footnote{More precisely, defining the embedding as $e:\mathcal{D}^k\to \mathbb{R}^N$, any background data sample $x\in\mathbb{R}^N$ in the absence of noise always belongs to the image of the embedding $\text{Im}(e)$. Therefore, for all background data points, there is a composition $f_\theta\circ e:\mathcal{D}^k\to \mathcal{M}^{k_l}$ of the embedding with the encoder map. With a slight misuse of notation, we define the encoder map as $f_\theta:\mathcal{D}^k\to\mathcal{M}^{k_l}$.} while $g_\phi:\mathcal{M}^{k_l}\to\mathcal{\tilde{D}}^k$ denote the decoder map with $\mathcal{M}^{k_l}$ denoting the latent manifold and $\mathcal{\tilde{D}}^k$ the output manifold. Perfect reconstruction of the background data manifold implies that it is identical to the output manifold  $\mathcal{\tilde{D}}^k=\mathcal{D}^k$. However, if the latent manifold $\mathcal{M}^{k_l}$ does not admit a global embedding of $\mathcal{D}^k$, the condition can never be satisfied even considering the universal approximation property of the encoder and decoder in the zero-noise limit. The phase space data of particle scattering and decay generally live on manifolds with non-trivial topologies (and therefore cannot be embedded assuming trivial latent topology with same dimensions), some of which are discussed in the next subsection.  

\begin{figure}[t!]
\centering
\includegraphics[scale=.35]{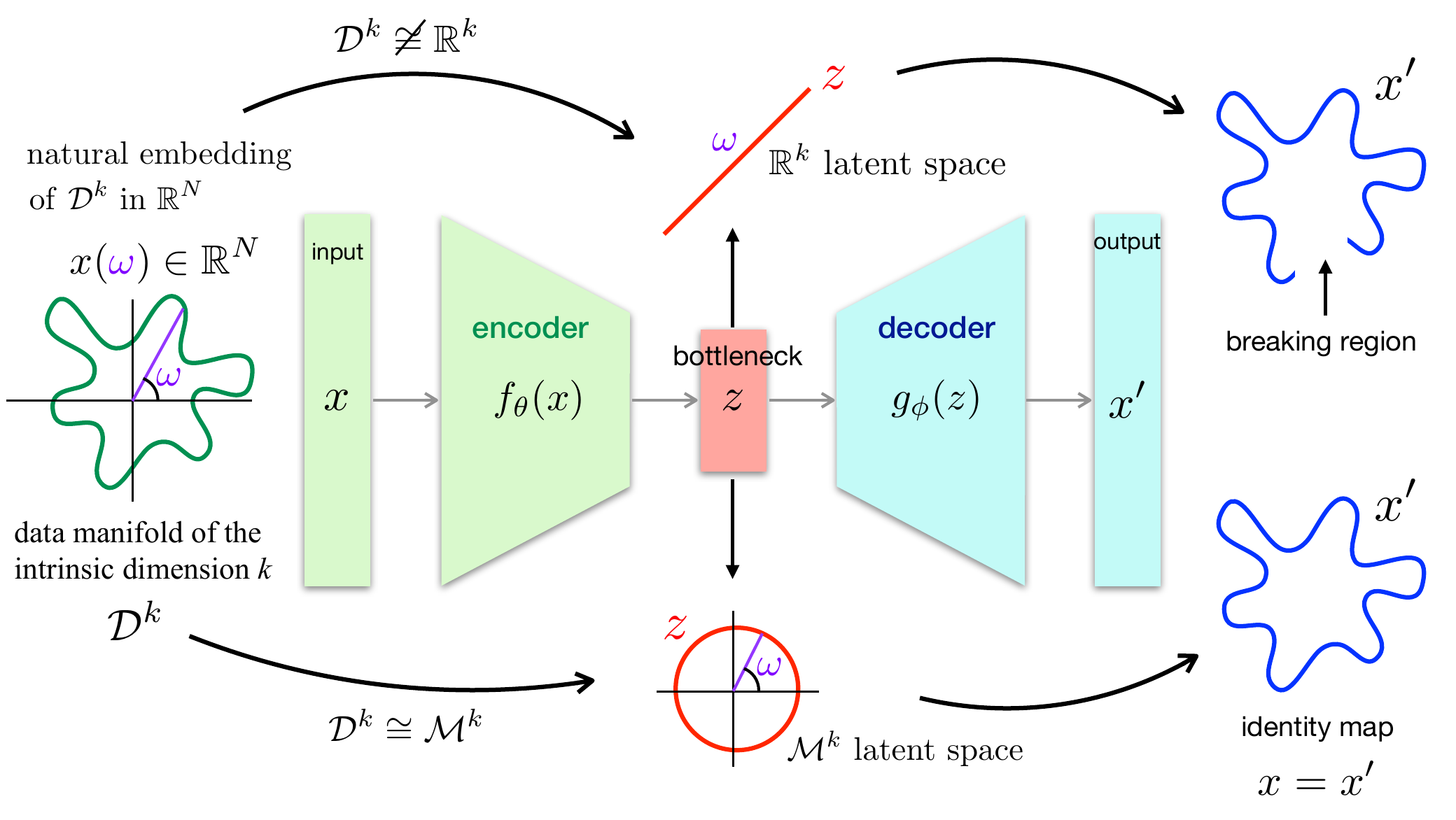}
\caption{Successful and unsuccessful global fits of data with different latent manifolds.   
\label{concept}
}
\end{figure}

If the topology of a background manifold is known a priori, one may construct a latent space such that the latent manifold $\mathcal{M}^k$ is homeomorphic to $\mathcal{D}^k$.
In this case, a global continuous map, $f_\theta: \mathcal{D}^k \mapsto \mathcal{M}^k$,
and its inverse, $g_\phi = f^{-1}_\theta: \mathcal{M}^k \mapsto \mathcal{D}^k$, can be realised by the encoder and decoder, respectively (see Fig.\ \ref{concept}). 
The background data is then minimally and faithfully described in the latent space, which, in principle, enhances the anomaly detection capability to its maximum.

\subsection{Topologies of momentum data} 

There are various cases in which the momentum data of final state particles exhibit non-trivial manifolds. 
For final states involving $n$ particles with given masses, those manifolds are naturally embedded in the $N$-dimensional Euclidean space, ${\mathbb R}^N$, with $N = 3 n$, where the coordinate of ${\mathbb R}^N$ is given by three-momenta of the particles, $(p^i_x, p^i_y, p^i_z)$ with $i = 1, \cdots, n$.    
Here we list some notable examples:
\begin{description}

\item[$S^{ 3n - 4 }$:~]

The momentum data of $n$ distinguishable particles with fixed total energy and momentum exhibits a data manifold holomorphic to $S^{ 3n - 4 }$ due to energy-momentum conservation \cite{Batson:2021agz}. 
For example, suppose a massive particle decays into three distinguishable particles, $X \to abc$, and the three momenta are measured at the rest frame of $X$. 
In that case, the final state momentum data lives on $S^5$.
Another example for the $S^5$ data manifold is the production of three distinguishable particles at a lepton collider with a fixed collision energy, e.g.\ $e^+ e^- \to \mu^+ \mu^- \gamma$.

\item[$\bigotimes_{i=1}^m S^{n_i}$:~]

Products of multiple hyperspheres arise when several productions and decays are combined. 
Consider, for example, the leptonic top decay, $t \to b W$ followed by $W \to \ell \nu$.
The initial decay, $t \to b W$, and the subsequent one, $W \to \ell \nu$, provide two $S^2$ corresponding to these two-body decays.   
Therefore, the momentum data of $(b,l,\nu)$ at the top rest frame exhibits $S^2 \otimes S^2$.
Next, consider the three particle production at a lepton collider, $e^+ e^- \to H Z \gamma$, followed by $H \to \tau^+ \tau^-$ and $Z \to \mu^+ \mu^-$. 
The production gives $S^5$, while each two-body decay provides $S^2$.  
The momentum data of the five final state particles, $(\gamma, \tau^+, \tau^- \mu^+, \mu^-)$, therefore exhibits $S^5 \otimes S^2 \otimes S^2$.

\item[$\mathds{RP}^2$:~]

Consider the production of a pair of identical particles $ab$.
When the momentum data is presented at the rest frame, ${\bf p}_a = - {\bf p}_b$ due to momentum conservation, and the identification $({\bf p}_a, {\bf p}_b) \sim ({\bf p}_b, {\bf p}_a)$ is implied since $a$ and $b$ are indistinguishable.  
This leads to the phase space manifold of the real projective plane, $\mathds{RP}^2$, an $S^2$ with antipodal points identified.  
The real projective plane may arise as a submanifold of the momentum data.
For instance, for the hadronic top decay, $t \to W b$, $W \to jj$, two jets from the $W$ decay are indistinguishable.\footnote{The two jets may be somewhat distinguishable by resorting to charm-tagging \cite{ATLAS:2022qxm,CMS-PAS-BTV-16-001}.}
Therefore, the momentum data of $(b, j, j)$
measured at the top rest frame exhibits $S^2 \otimes \mathds{RP}^2$.

\end{description}

\section{Construction of topologically non-trivial latent spaces} 
\label{sec:top_embed}
Before discussing ways to induce topologically non-trivial latent spaces, one should first consider the compositional nature of deep neural networks where, in general, there would be multiple hidden representations before and after the latent layer of an autoencoder. Even if the ideal input data manifold is known, usual gradient-based training does not explicitly control these constituent maps to follow a particular geometry in $\mathbb{R}^N$ for noisy data. For most non-trivial manifolds encountered in particle physics, there is a minimum Euclidean dimension $k_0$, where there exists a global embedding of the manifold guaranteed by various embedding theorems~\cite{Munkres2000Topology,c992e7fe-9a8a-3576-849e-c8a0d8dab89c,38b777be-fb0a-3998-b108-baf022f5dfd4}. Therefore, in densely connected networks which assume trivial $\mathbb{R}^{n_h}$ topology in each $n_h$-th hidden representation, it is necessary that $k_0 \leq 
 n_h$. However, this alone is not sufficient.
 Since our main focus is on clearing topological obstructions, we construct \emph{latent layers that can propagate a global embedding} of the known data manifold to the decoder. In other words, the latent layer and the whole autoencoder fulfil the necessary conditions for the propagation of a global embedding of the known background data manifold in the output space. As we shall see in Section~\ref{sec:toy_clear}, experiments on toy datasets show that such conditions are enough for an autoencoder to learn a global embedding of non-trivial manifolds.

For a background manifold  $\mathcal{D}^{k}$ of intrinsic dimensions $k$, let $k_0$ be the minimum Euclidean dimensions where $\mathcal{D}^k$ can be faithfully embedded. Let $k_l\geq k_0$ be the minimum dimensions (generally at the latent layer) of any hidden representation, all assuming the trivial topology. Even though this fulfils the necessary conditions, it can lead to inefficient anomaly detection for any signal manifold $\mathcal{D}_S^{k'}$, with $k<k'\leq k_l$. 
This is because there will be well-behaved local charts of $\mathcal{M}_S^{k'}$ in $\mathbb{R}^{k_l}$ for $k'\leq k_l$ even if there is no global embedding. On the other hand, if one fixes $k_l < k'$ for the dimension of the latent representation, it restricts the existence of local charts for $\mathcal{M}^{k'}_S$. 
Therefore, we construct the latent space manifold $\mathcal{M}^{k_l}$ with $k \leq k_l < k'$ while requiring it 
to admit a global embedding of the data manifold $\mathcal{D}^k$.

To realise such topologically non-trivial latent spaces, we construct the bottleneck part of the autoencoder using two layers.  
We call the first and second layers the ``preparation layer'' and ``latent layer'', respectively. 
The value represented by the $i$-th neuron of the preparation layer is denoted by $y_i$, while $z_i$ denotes the value represented by the $i$-th neutron of the latent layer.
The realisation of a non-trivial latent manifold, ${\cal M}^k$, in the latent layer with $n$ neurons, is essentially the same as the embedding of ${\cal M}^k$ into ${\mathbb R}^n$.
Here, we describe concrete examples:
\begin{description}

\item[$S^n$:~]
$n+1$ neurons are given to both the preparation and latent layers. 
The neuron values in the latent layer are assigned from those of the preparation layer as $z_i = y_i / r$ with $r \equiv \sum_{i=1}^{n+1} y_i^2$.

\item[$\bigotimes_{i=1}^m S^{n_i}$:~]
The latent space with this topology can be constructed by repeating the above construction multiple times. 
For example, to construct $S^{n_1} \times S^{n_2}$, we give $n_1 + n_2 + 2$ neurons both to the preparation and latent layers.
We construct $S^{n_1}$ from the first $n_1 + 1$ neurons by
$z_i = y_i/r_1$ ($i=1,\cdots,n+1$) with $r^2_1 = \sum_{i=1}^{n_1 + 1} y_i^2$
and 
$S^{n_2}$ from the remaining $n_2 + 1$ neurons by
$z_j = y_j/r_2$ ($j=n_1+1,\cdots,n_1 + n_2+2$) with $r^2_2 = \sum_{j=n_1 + 1}^{n_1 + n_2 + 2} y_j^2$.

\item[$\mathds{RP}^2$:~]
The minimum Euclidean dimension to embed $\mathds{RP}^2$ is 4.
We use the well-known embedding of $\mathds{RP}^2 \to {\mathbb R}^4$.
Three neurons are given to the preparation layer, while we give four neurons to the latent layer. 
The values of latent neurons are constructed as
$(z_1, z_2, z_3, z_4) = ( \tilde y_1^2 - \tilde y_2^2, \,\tilde y_1 \tilde y_2, \,\tilde y_2 \tilde y_3, \,\tilde y_3 \tilde y_1)$
with $\tilde y_i \equiv y_i / r$ and $r \equiv \sum_{i=1}^3 y_i^2$.
Note that $\tilde y_i$ represents $S^2$ as $\tilde y_1^2 + \tilde y_2^2 + \tilde y_3^2 = 1$.
Also, $(y_1, y_2, y_3)$ and $-(y_1, y_2, y_3)$ map to the same point in ${\mathbb R}^4$. 
The map is differentiable everywhere except for $y_1 = y_2 = y_3 = 0$.

\end{description}

 \section{Clearing Topological Obstructions on Toy datasets}
 \label{sec:toy_clear} 
In this section, we numerically evaluate the reconstruction of non-trivial manifolds with and without assuming non-trivial latent topology. To showcase the difficulty in reconstructing the input manifold of intrinsic dimensions $k$ via the trivial $\mathbb{R}^{k_l}$ topology and $k_l\leq k$ in the latent space, we consider a deep and symmetric autoencoder with encoder node dimensions 1024, 512, 256, 128, and 64 before the latent output layer.   All hidden layers have \texttt{ReLU} activation while the latent layer has \texttt{tanh} activation function.  
To see a recovery of global fitting with $\mathbb{R}^{k_l}$ latent spaces with higher $k_l$,
we additionally consider architectures with $k_l\geq k_0$, where $k_0$ is the minimum Euclidean dimensions where there are global embeddings of the data manifold. Additional details of the training can be found in Section~\ref{sec:net_arch_train}, with the difference being that all inferences are made on a single randomly initialised network trained for a maximum of one thousand epochs.

\subsection{2-sphere embedded in ${\mathbb R}^3$}
We first take a relatively straightforward example of $S^{2}$ embedded in $\mathbb{R}^3$ as a non-trivial two-dimensional data manifold. 500k coordinate samples $(x,y,z)$ with $x^2+y^2+z^2=1$ in $\mathbb{R}^3$ are sampled with a seeded random number generator which is divided into 400k training and 100k validation samples. Uniform distribution on $S^2$ was achieved by uniformly sampling $\phi$ and $\cos{\theta}$, where $\phi$ and $\theta$ are the azimuthal and polar angles of the spherical coordinate system, respectively.  For the $S^2$ latent-topology autoencoder ($S^2$-AE), the encoder and the decoder have a single hidden layer of sixty-four nodes with extrinsic latent space dimensions (the number of neurons in the latent layer) of three. 
As described in the previous section, the degrees of freedom of the latent output are reduced to reflect the induced $S^2$ topology.  
For trivial latent spaces $\mathbb{R}^k$, the value of $k$ fixes the encoder output (and therefore the decoder input) dimensions.  We will denote such autoencoders as $\mathbb{R}^k$-AE. 

\begin{figure}[t!]
	\centering
	\includegraphics[scale=0.2]{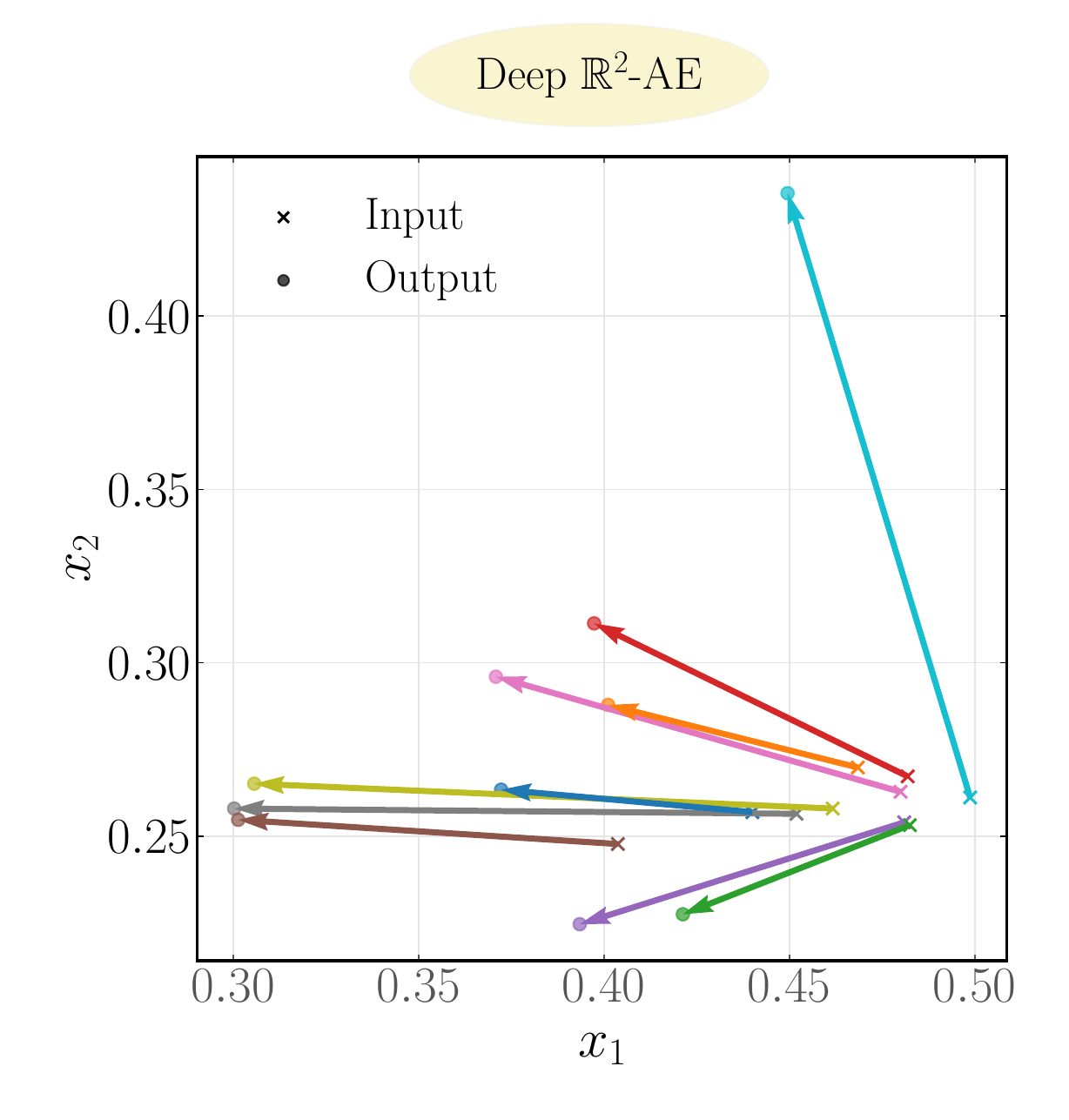}
    \includegraphics[scale=0.2]{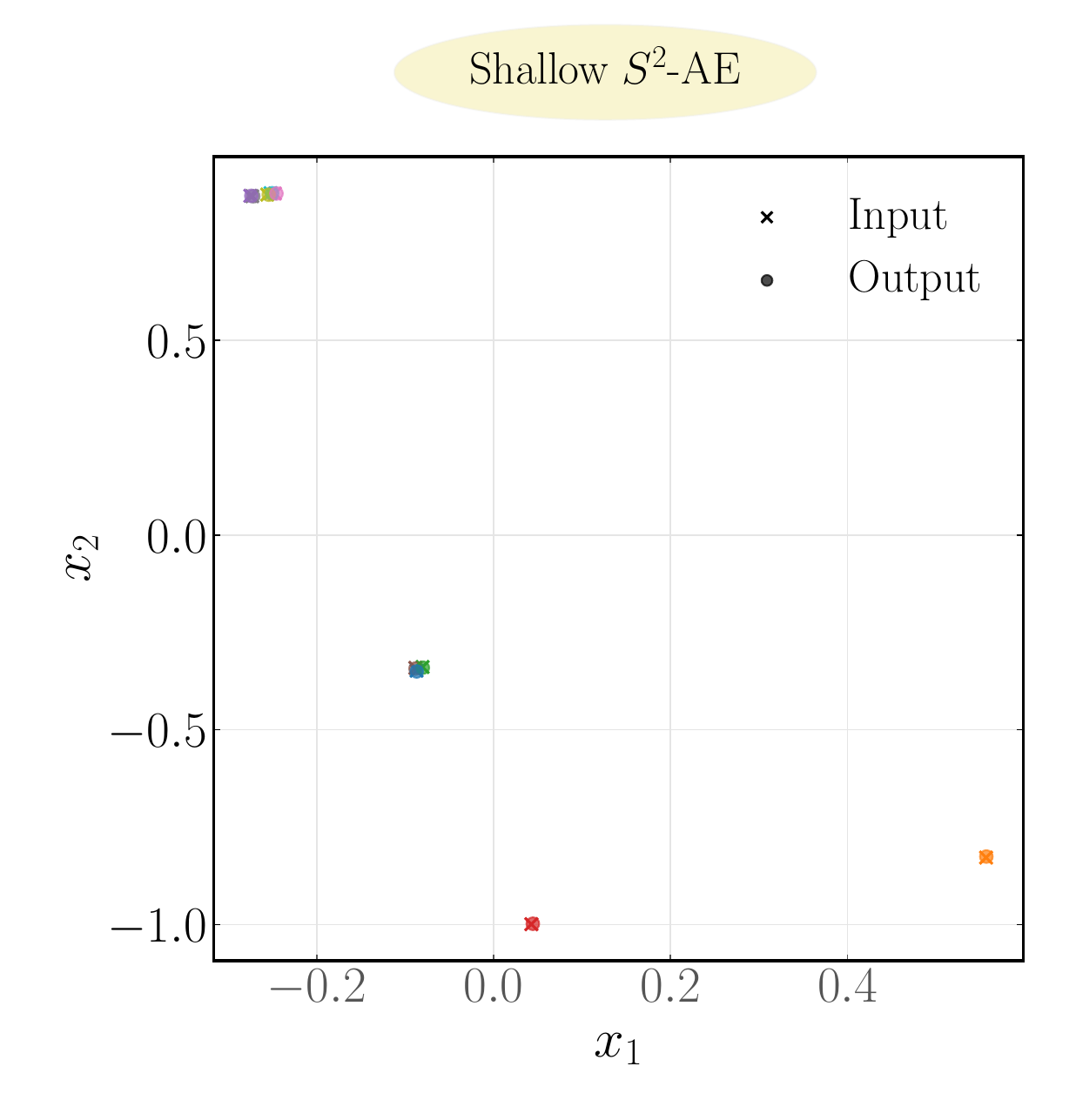}
    \includegraphics[scale=0.2]{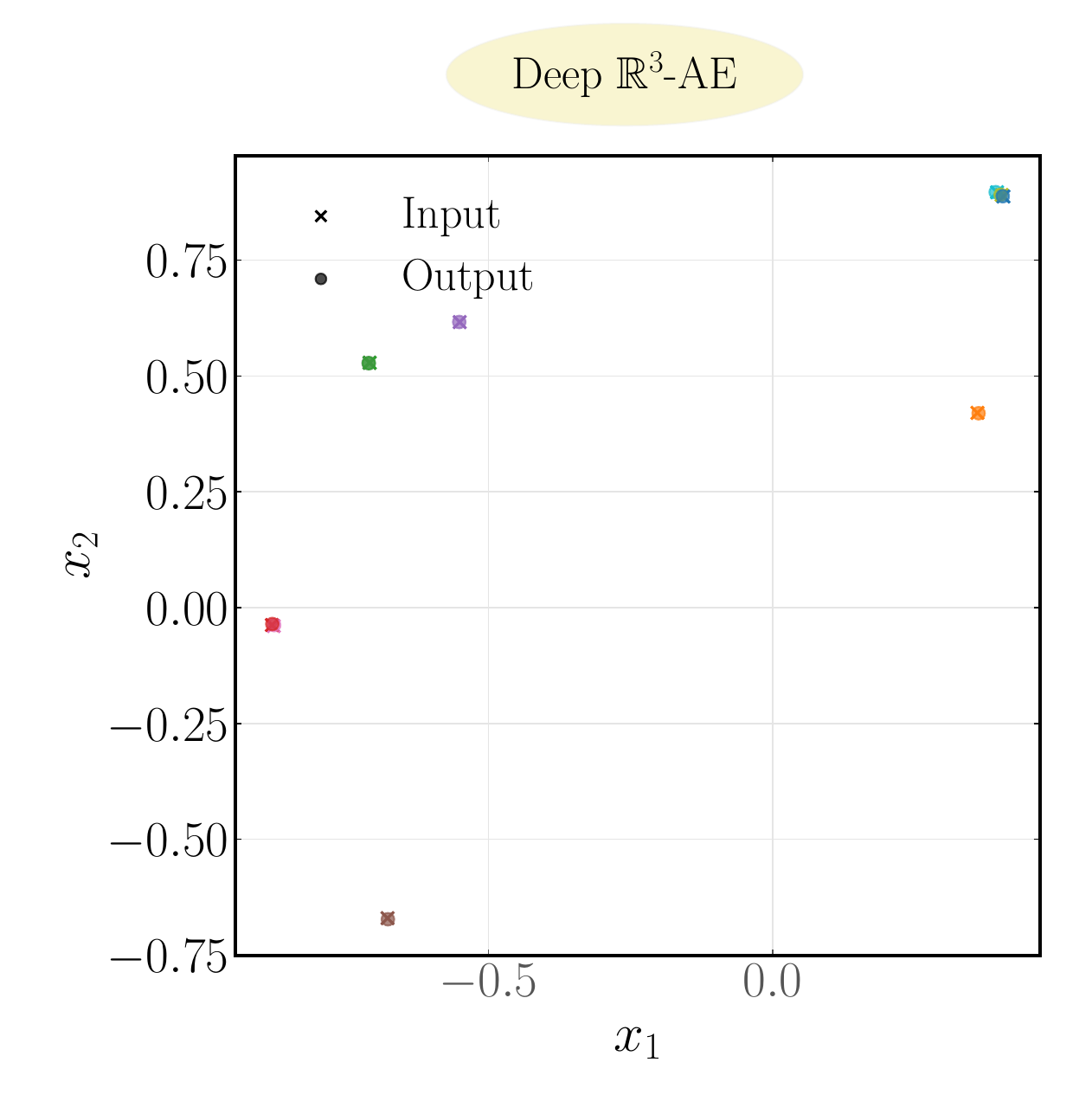}\\
    \includegraphics[scale=0.2]{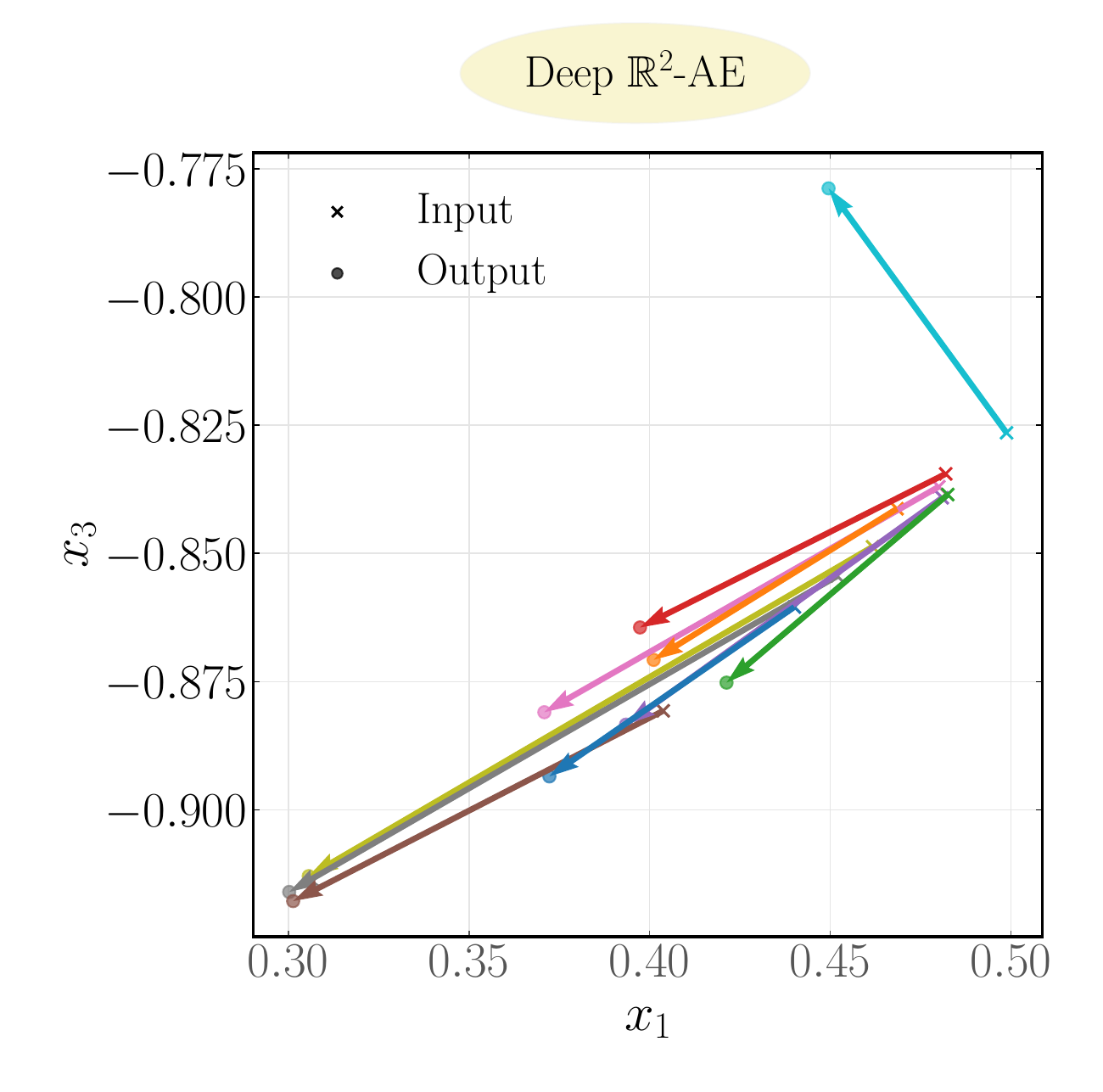}
    \includegraphics[scale=0.2]{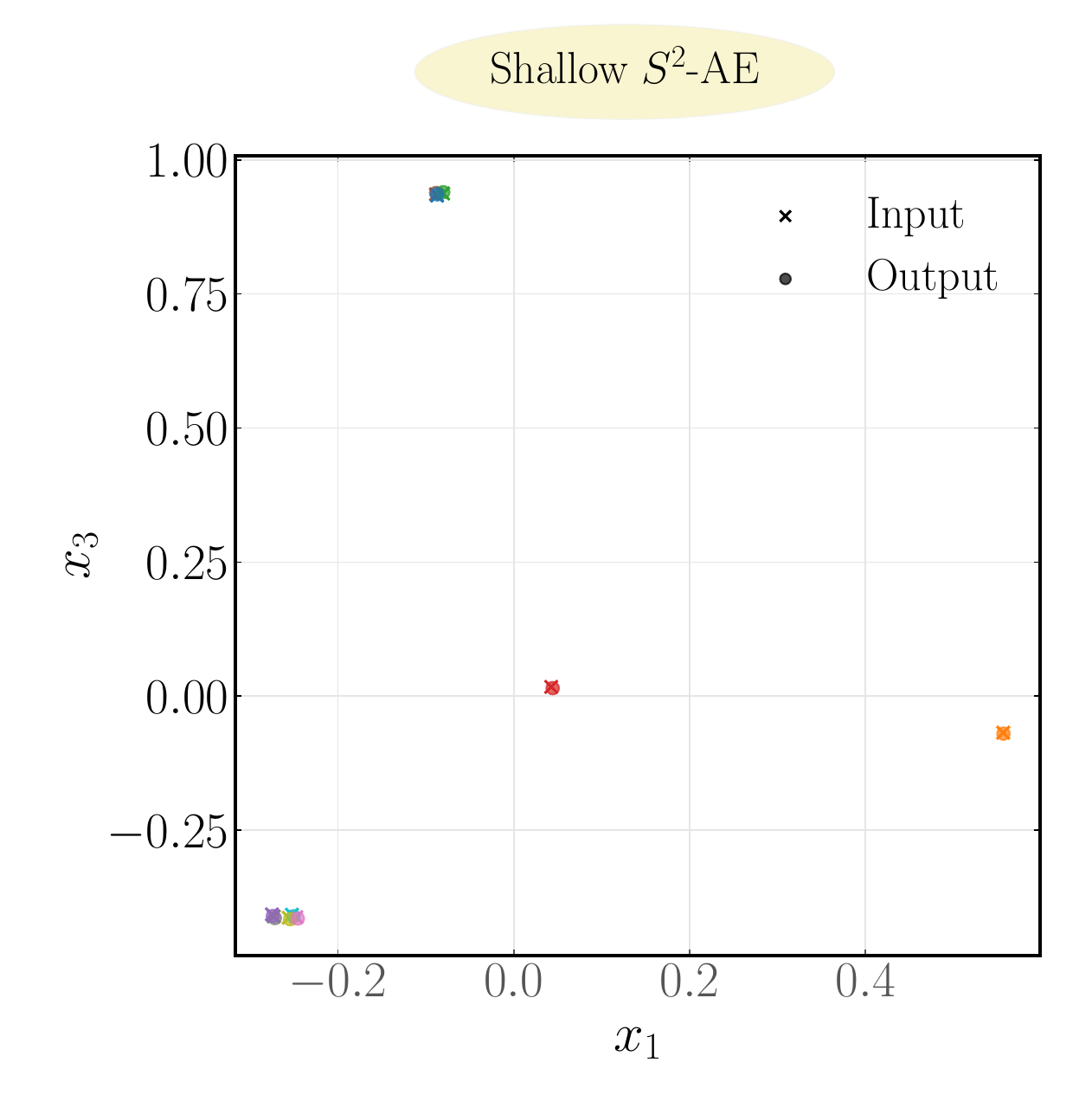}
    \includegraphics[scale=0.2]{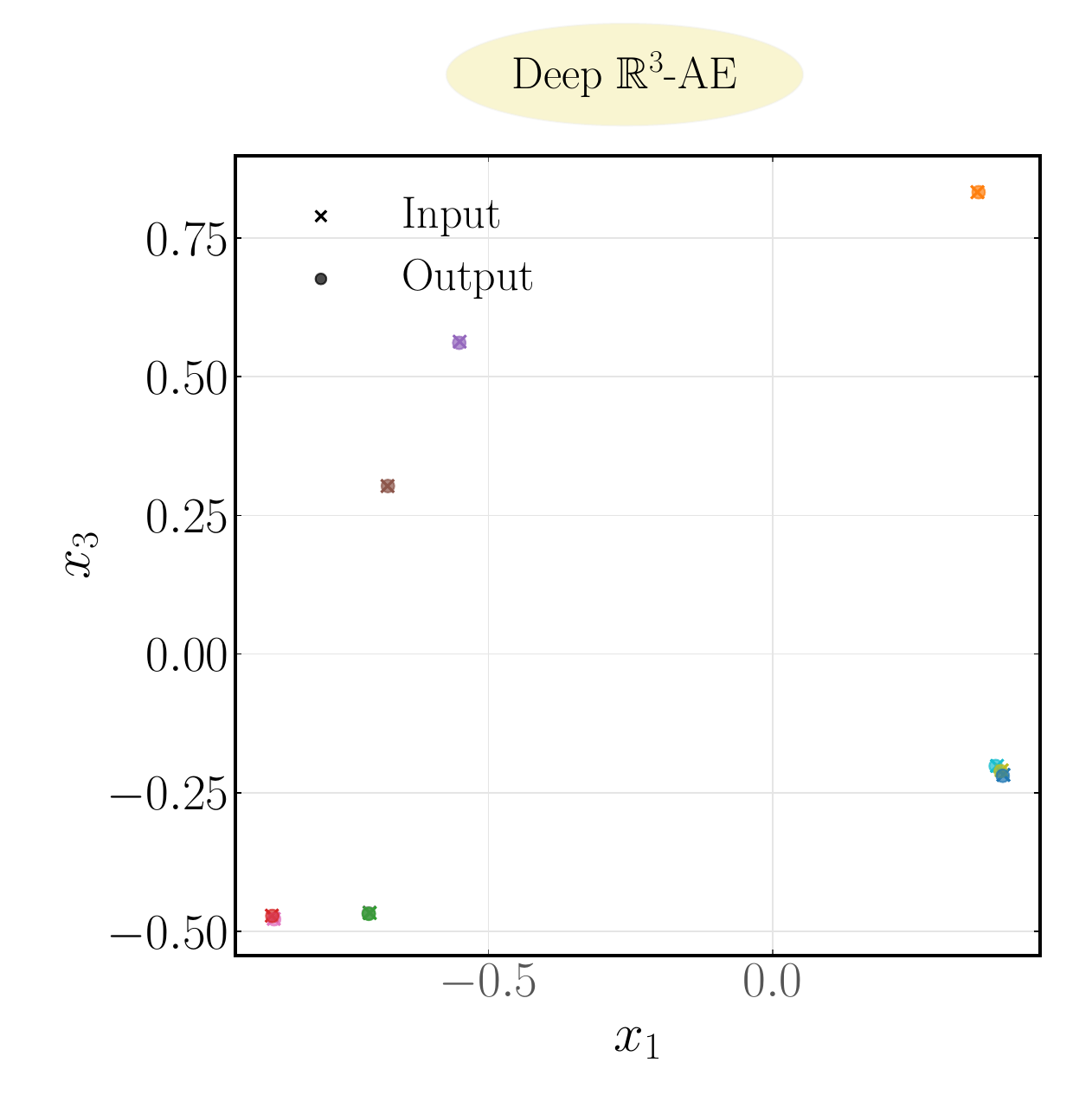}\\
    \includegraphics[scale=0.2]{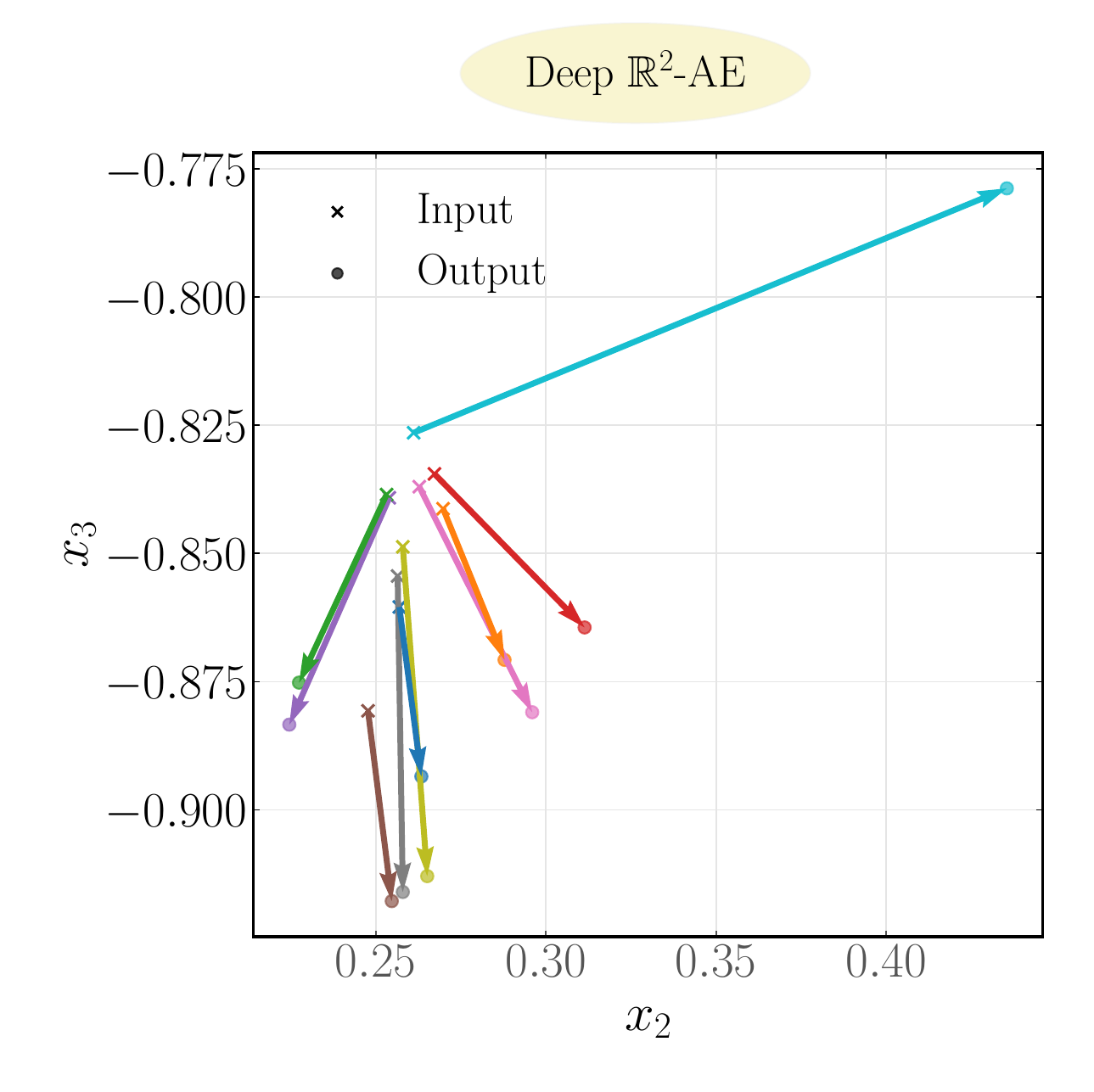}
    \includegraphics[scale=0.2]{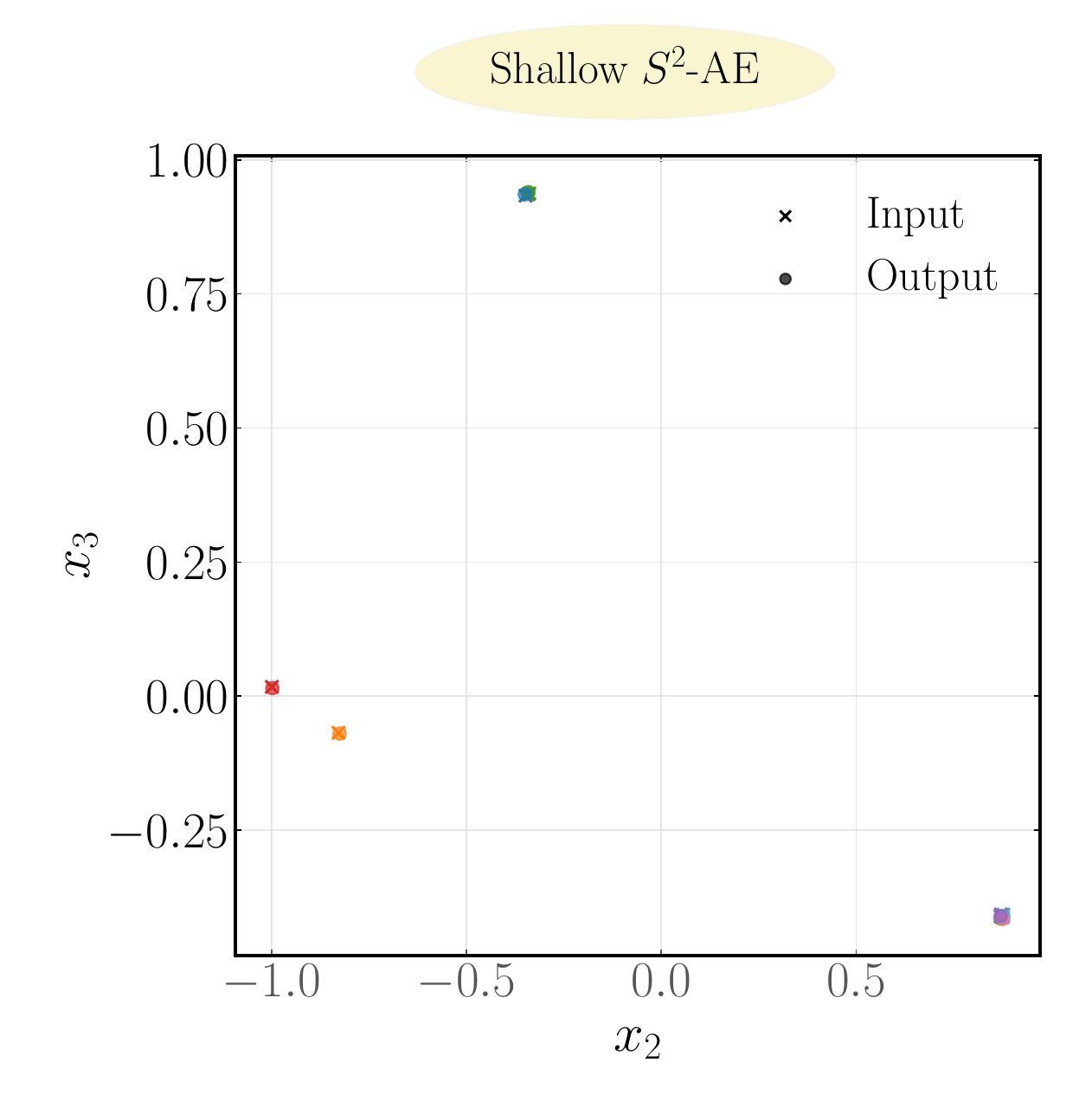}
        \includegraphics[scale=0.2]{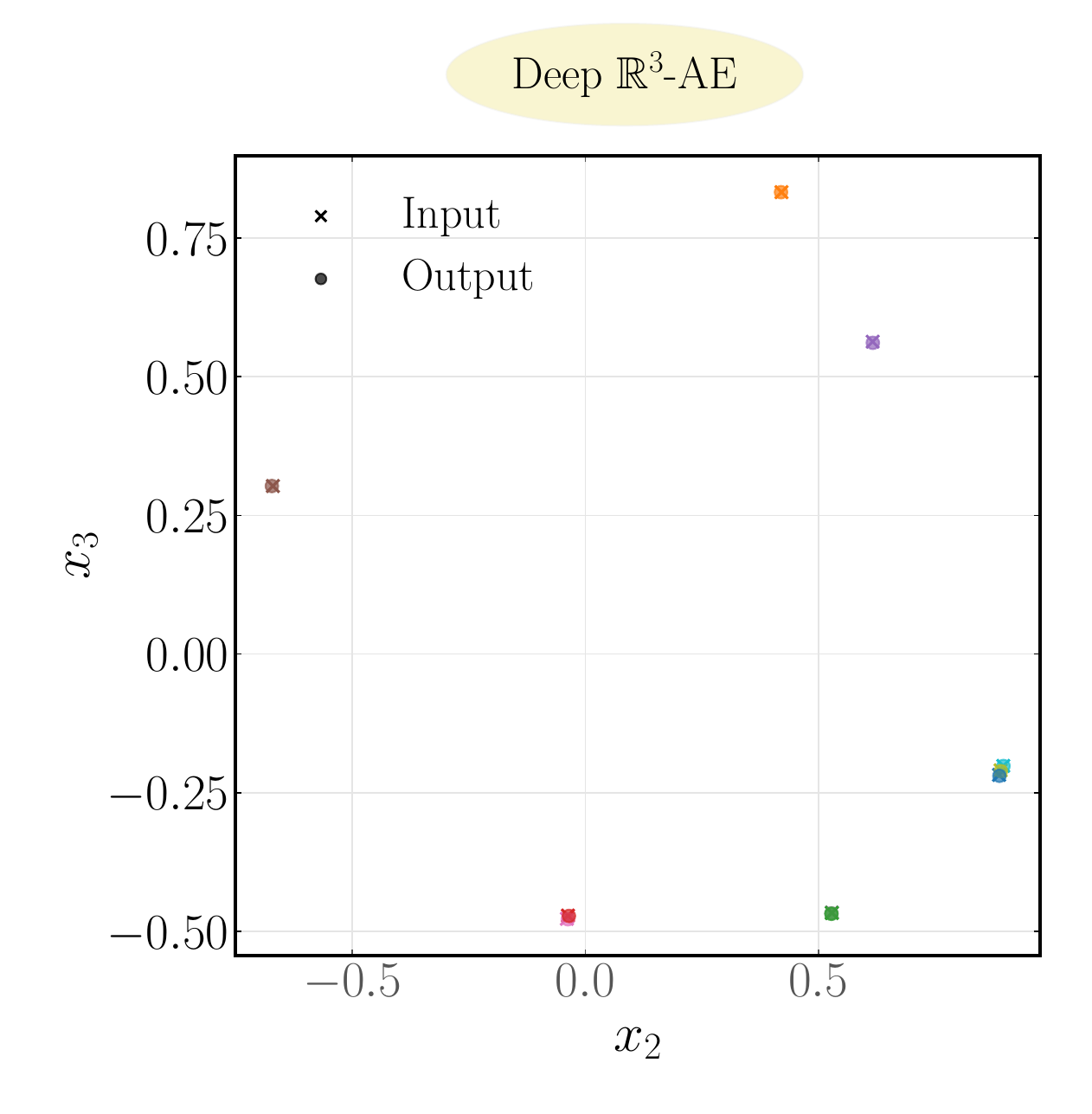}
	\caption{Input--output pairs for ten samples with highest reconstruction error for different latent spaces. The dataset is $S^2$ embedded in $\mathbb{R}^3$. 
    The 3D data is projected onto three different 2D planes. }
	\label{fig:s2_worst_point_scatter}
\end{figure}

For each converged network, a 2D projected scatter plot of the input and the output of ten samples with the highest reconstruction error within the validation data are shown in Fig.~\ref{fig:s2_worst_point_scatter}. For each coloured cross denoting an input sample, circles of the same colour denote the sample's output.   
Clearly, $\mathbb{R}^2$-AE cannot fit all points on the sphere with very large discrepancies between the inputs and the outputs.  On the other hand, both $\mathbb{R}^3$-AE and $S^2$-AE can fit the data over the whole 2-sphere since the input and the output of the worst points completely overlap.

\begin{figure}[t!]
	\centering
	\includegraphics[scale=0.15]{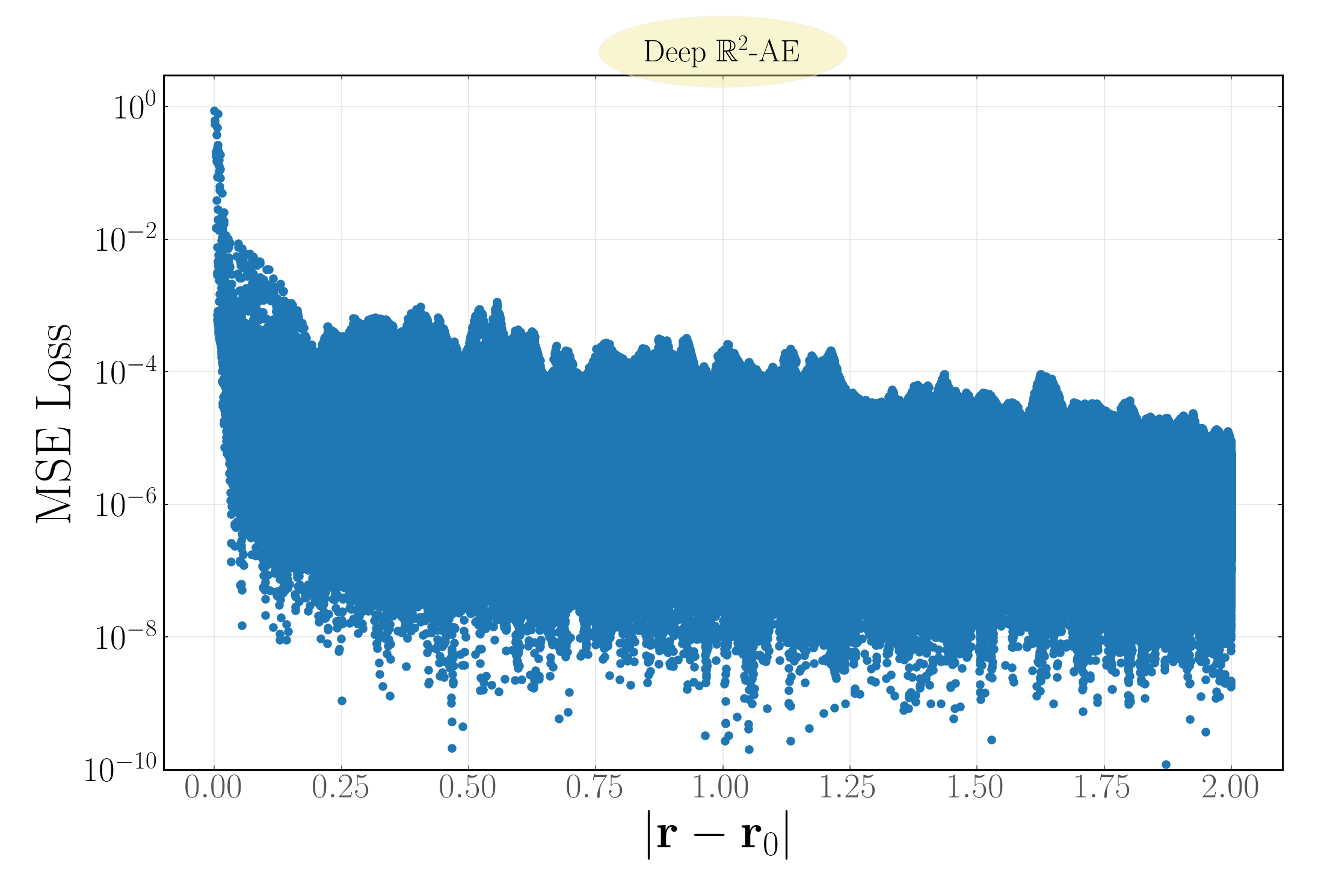}
    \includegraphics[scale=0.15]{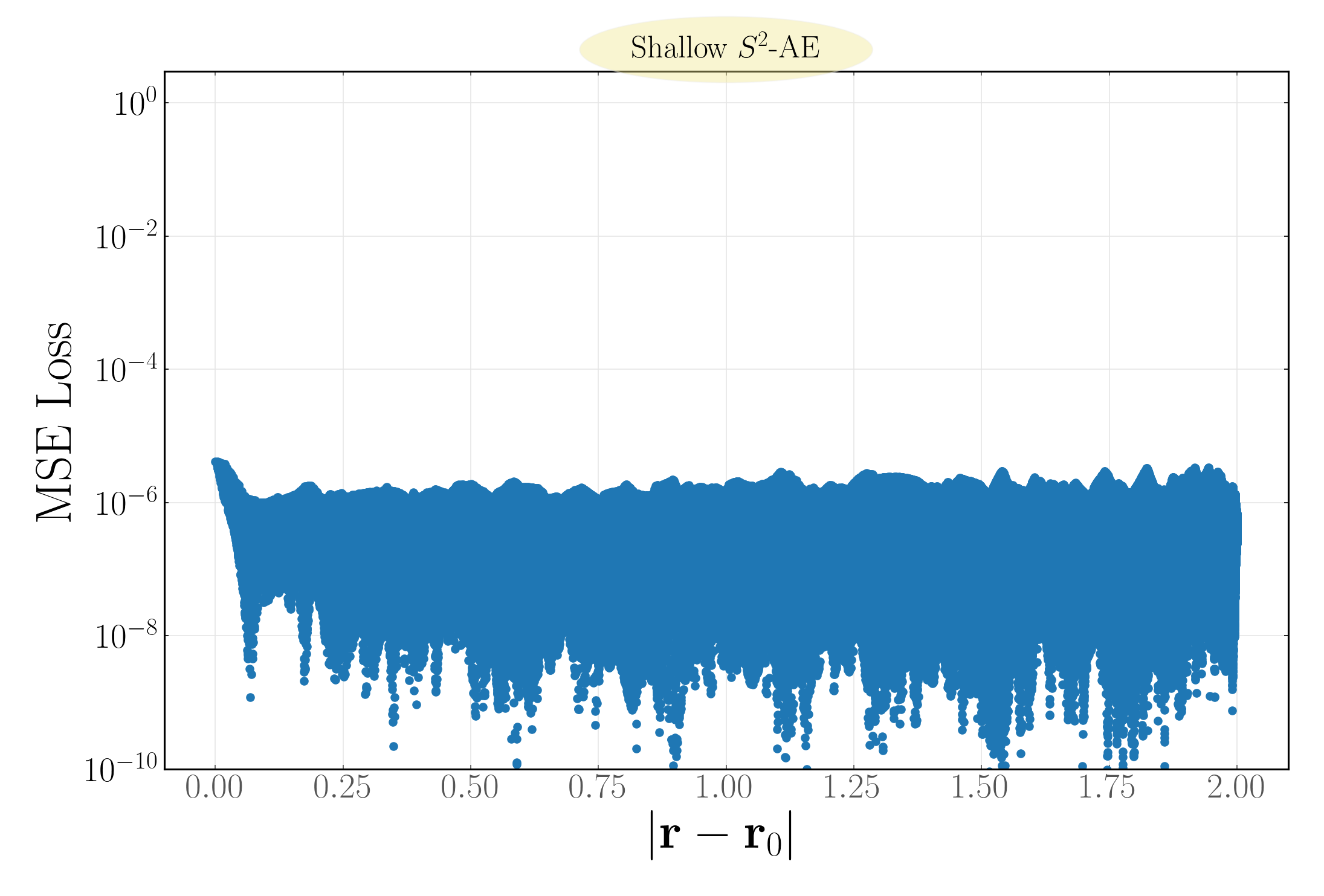}
    \includegraphics[scale=0.15]{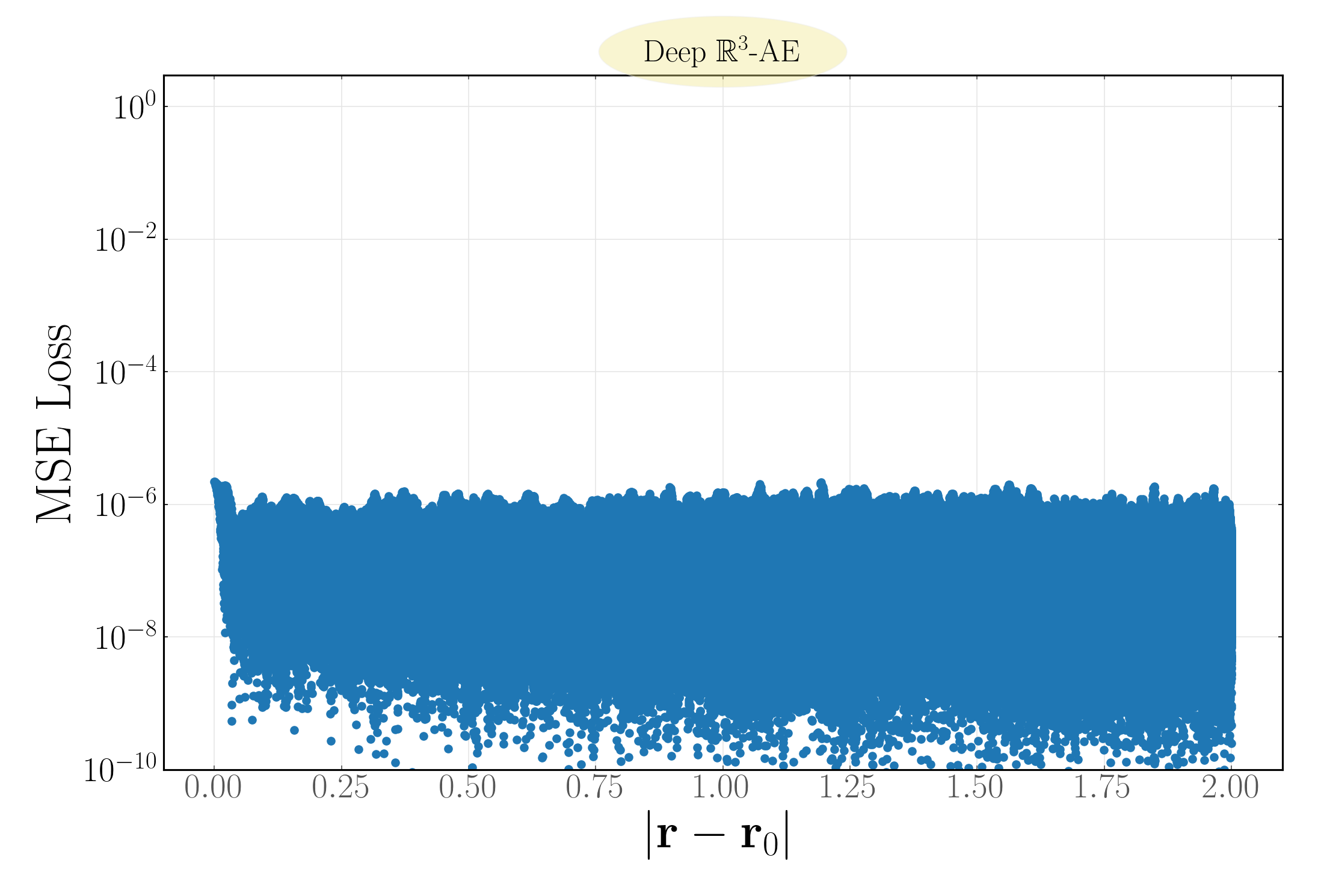}
    \caption{Loss-versus-distance plots for three latent layers. For each sample $\mathbf{r}$, the distance $|\mathbf{r}-\mathbf{r}_0|$ is measured from the sample $\mathbf{r}_0$ with the highest loss. Topological anomaly is visible as a peak on the plot for $\mathbb{R}^2$ latent space. Two million data points were used to properly visualise topological anomalies.}
	\label{fig:lossvsdist_s2data}
\end{figure}

We may further analyse the global goodness of fit by plotting loss-versus-distance shown in Fig.~\ref{fig:lossvsdist_s2data}. 
A loss-versus-distance plot is a helpful tool introduced in~\cite{Batson:2021agz}, visualizing the loss of samples versus their Euclidean distance from the point, $x_0$, with the largest loss in the dataset. 
Consider, for example, a relation between $S^n$ and ${\mathbb R}^n$.
If one removes a single point from $S^n$, the resulting space is homeomorphic to $\mathbb{R}^n$. 
This implies that $\mathbb{R}^n$-AE should be able to accurately fit $S^n$ data everywhere except for the neighbourhood of a point located somewhere on the $S^n$. 
We therefore expect that the loss values are high in the vicinity of the largest error point, $x_0$, and low for points distanced from $x_0$. 
This is exactly what is observed in the left plot of Fig.~\ref{fig:lossvsdist_s2data}, where the loss-versus-distance is shown for the $\mathbb{R}^2$-AE.
This behaviour is not visible in the other two plots for the $S^2$-AE (middle) and ${\mathbb R}^2$-AE (right), where their latent spaces allow for good global fit to $S^2$ data.
For more complicated manifolds, such as $\bigotimes_{i=1}^m S^{n_i}$ and $\mathds{RP}^2$, removing a single point does not yield a space homeomorphic to Euclidean spaces. 
For those data manifolds, we anticipate different behaviours of the loss-versus-distance plot.

The corresponding plots in Fig.~\ref{fig:s2_worst_point_scatter} and \ref{fig:lossvsdist_s2data} demonstrate that the $\mathbb{R}^3$-AE ($S^2$-AE) can find a global embedding (mapping) of the $S^2$ data manifold into its latent space. 
From the perspective of anomaly detection, while $\mathbb{R}^3$-AE can indeed fit the ``background'' $S^2$ manifold, $\mathbb{R}^3$  has higher intrinsic dimensions than the background manifold, which could result in poor anomaly detection capabilities for signals with intrinsic dimensions less than or equal to 3. 
For the $S^2$-AE case, the intrinsic data dimension is equal to the latent dimension, suggesting higher anomaly detection capabilities. 
This is explicitly checked in the numerical experiments with realistic collider data, even when the input space has much higher extrinsic dimensions than $k$.  
In any case, it is evident that $S^2$-AE effectively clears the topological obstructions for the $S^2$ background data by keeping the intrinsic latent dimensions no more than those of the background data manifold. 

{
\begin{figure}[t!]
	\centering
	\includegraphics[scale=0.2]{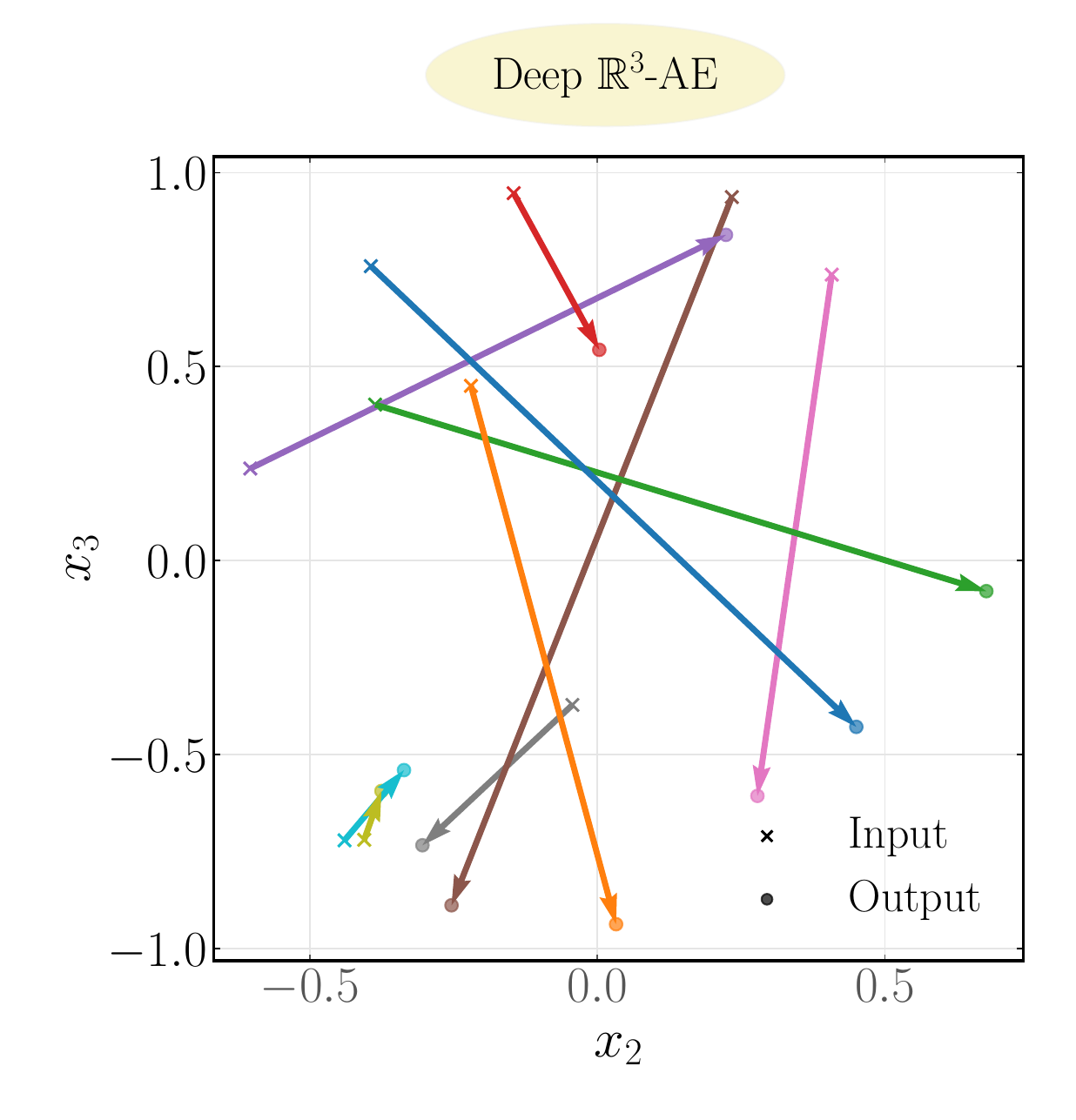}
    \includegraphics[scale=0.2]{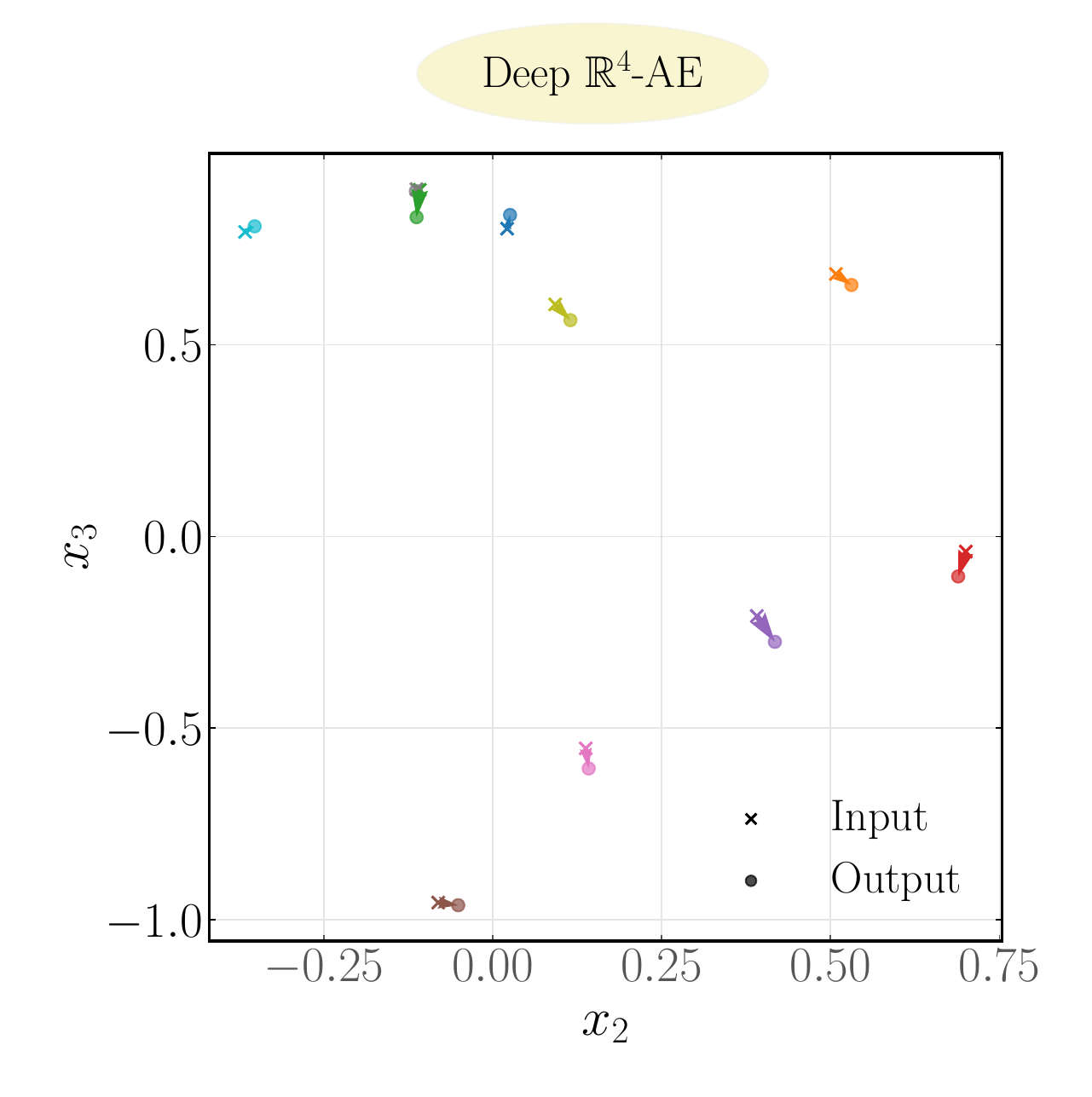}
    \includegraphics[scale=0.2]{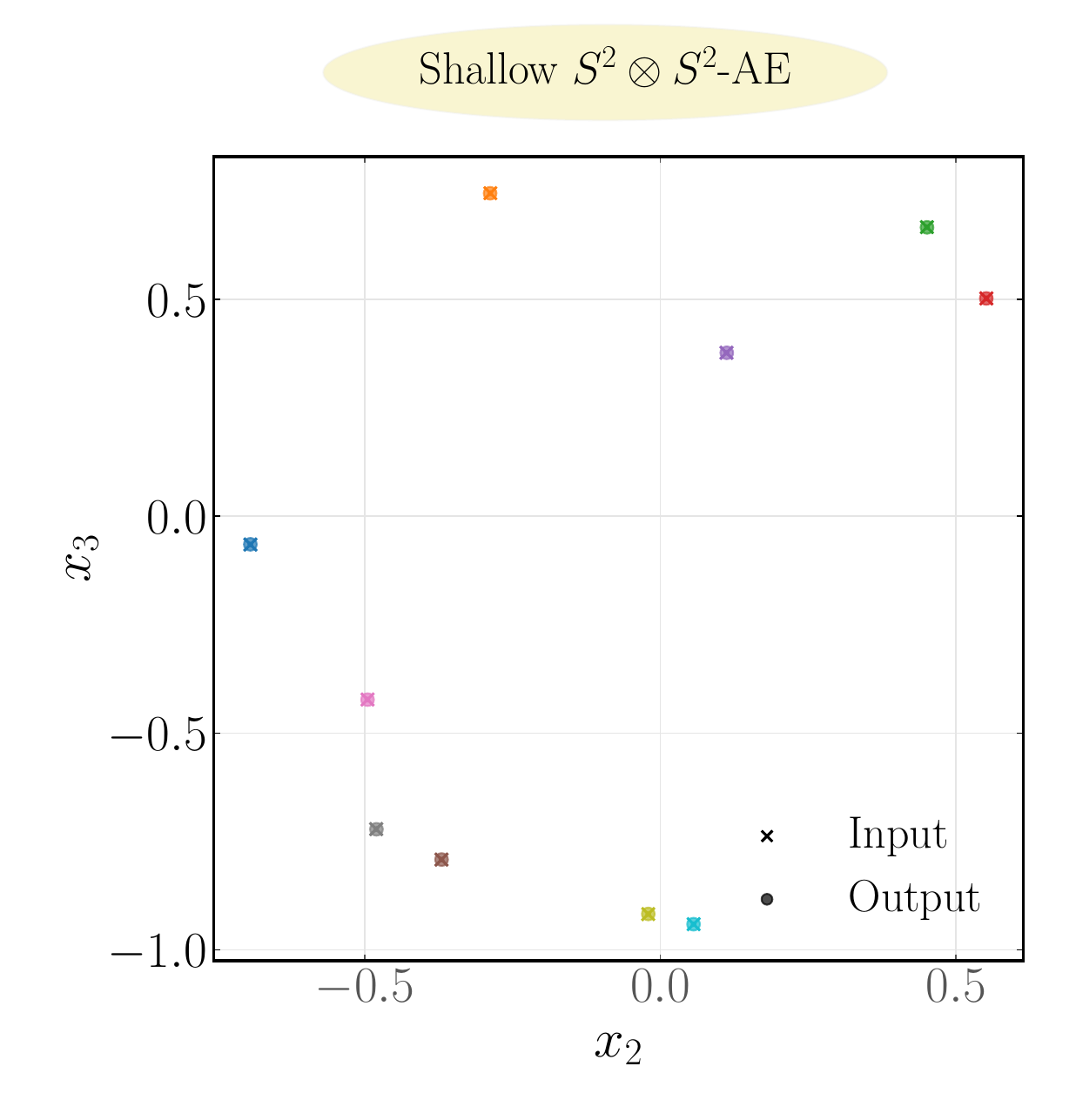}
    \includegraphics[scale=0.2]{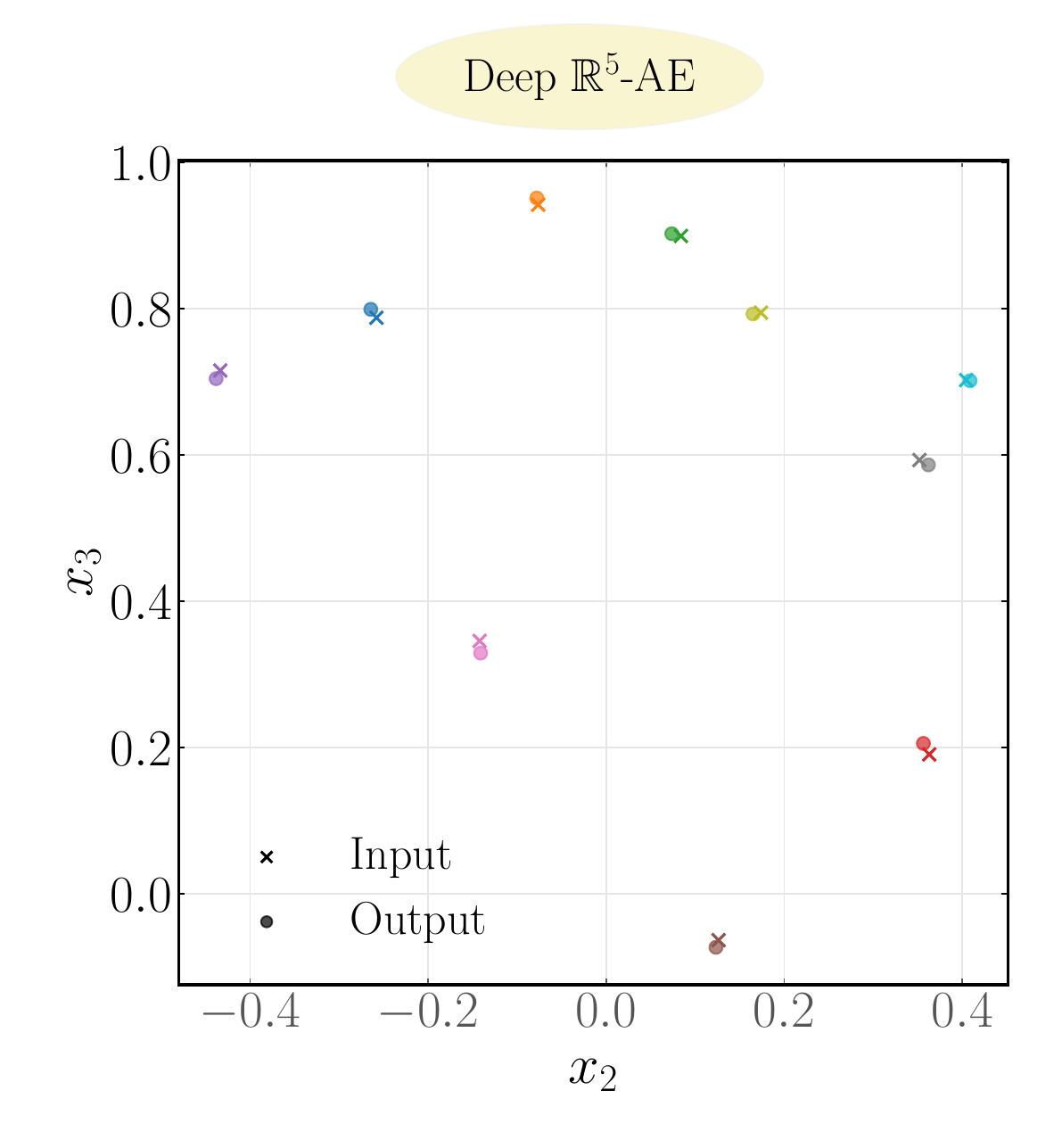}\\
    \includegraphics[scale=0.2]{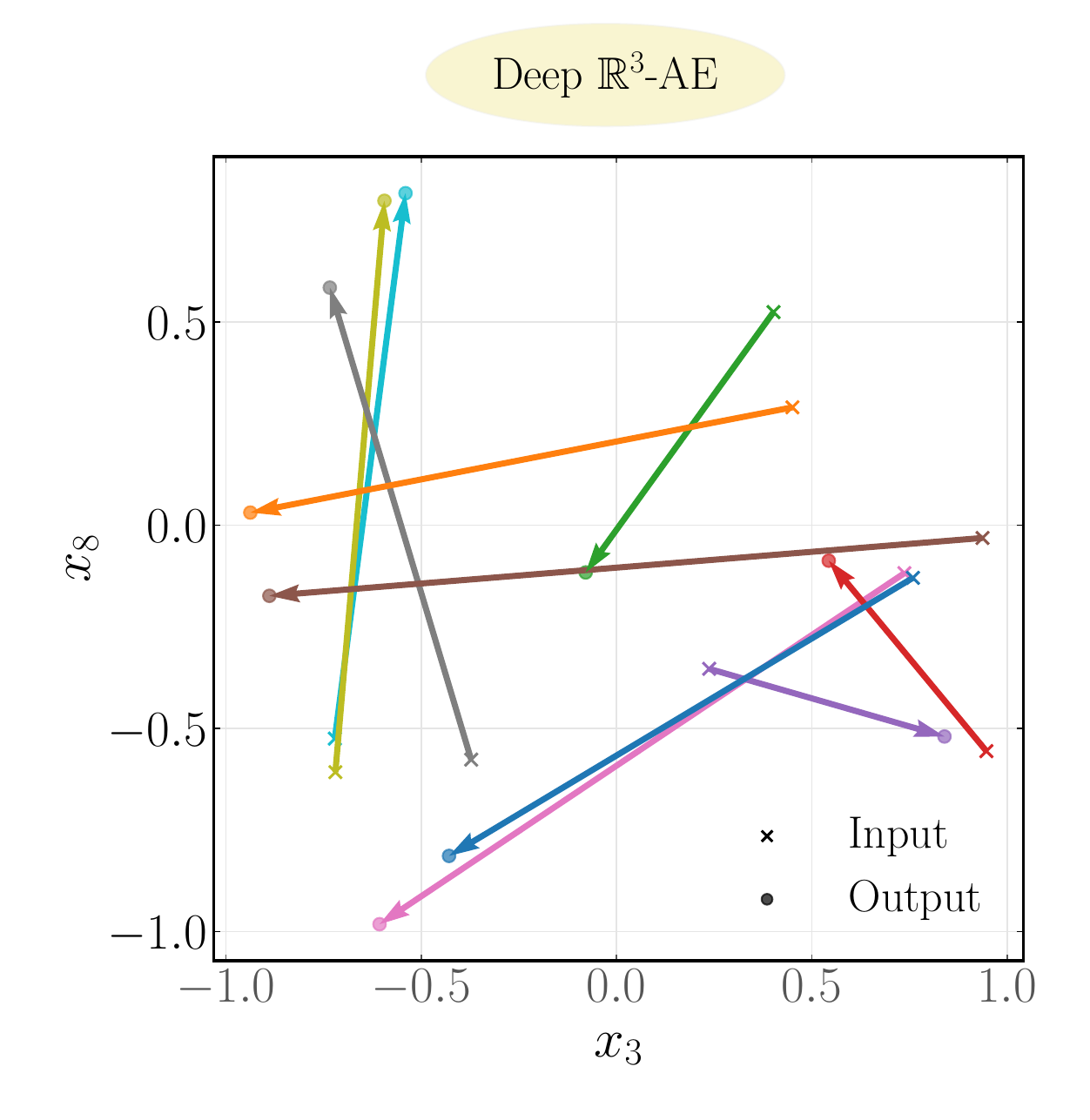}
    \includegraphics[scale=0.2]{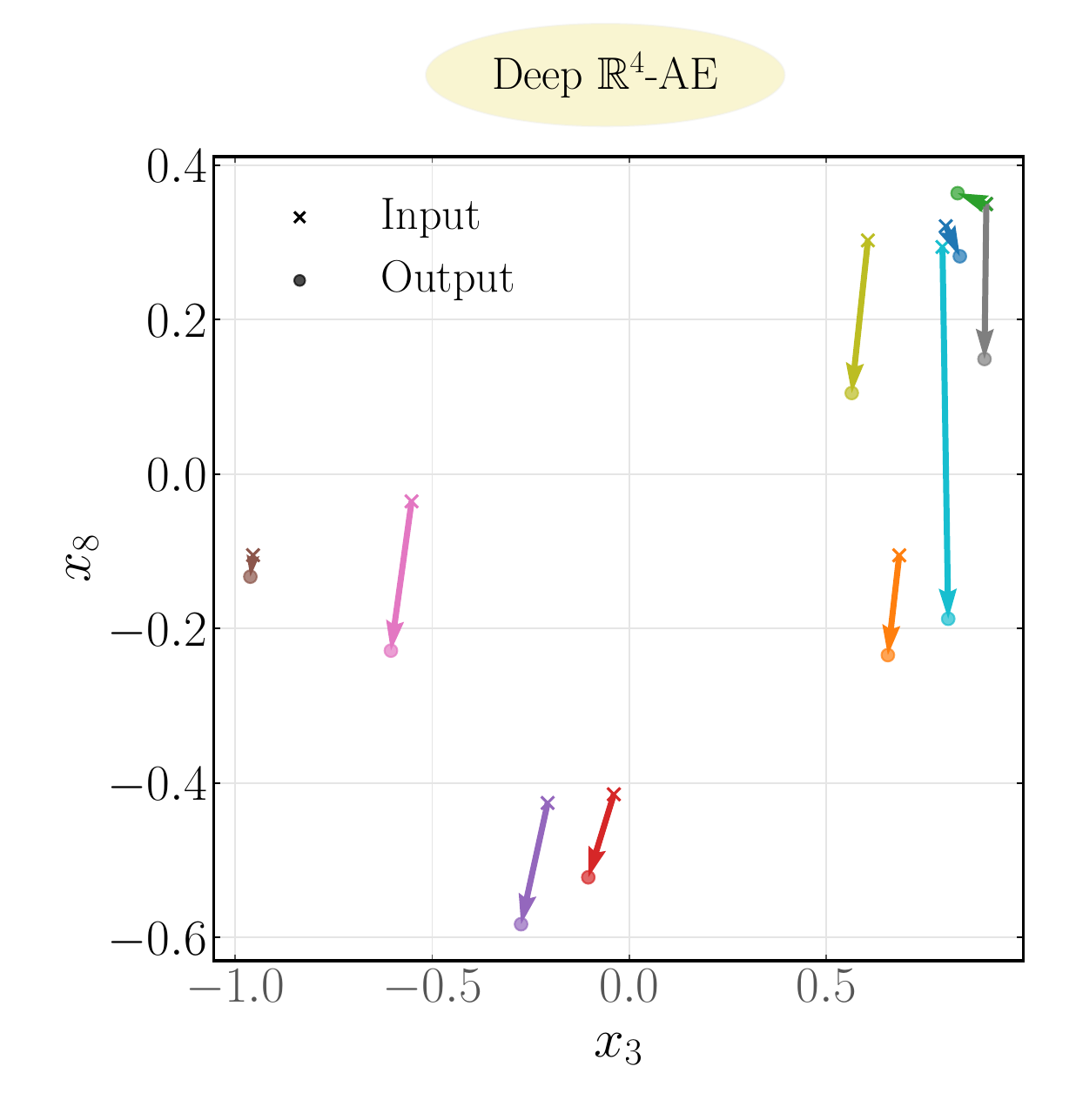}
    \includegraphics[scale=0.2]{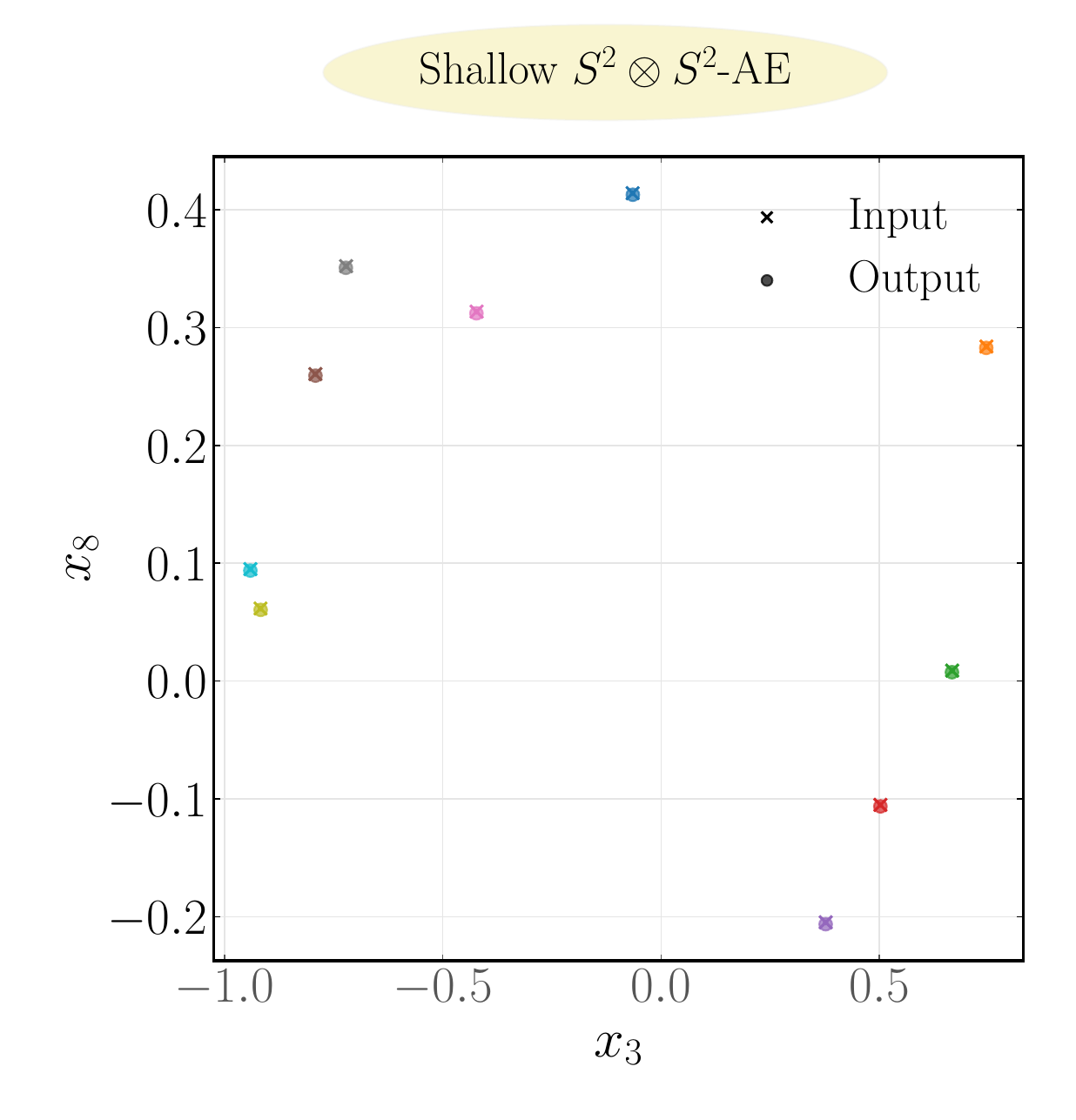}
    \includegraphics[scale=0.2]{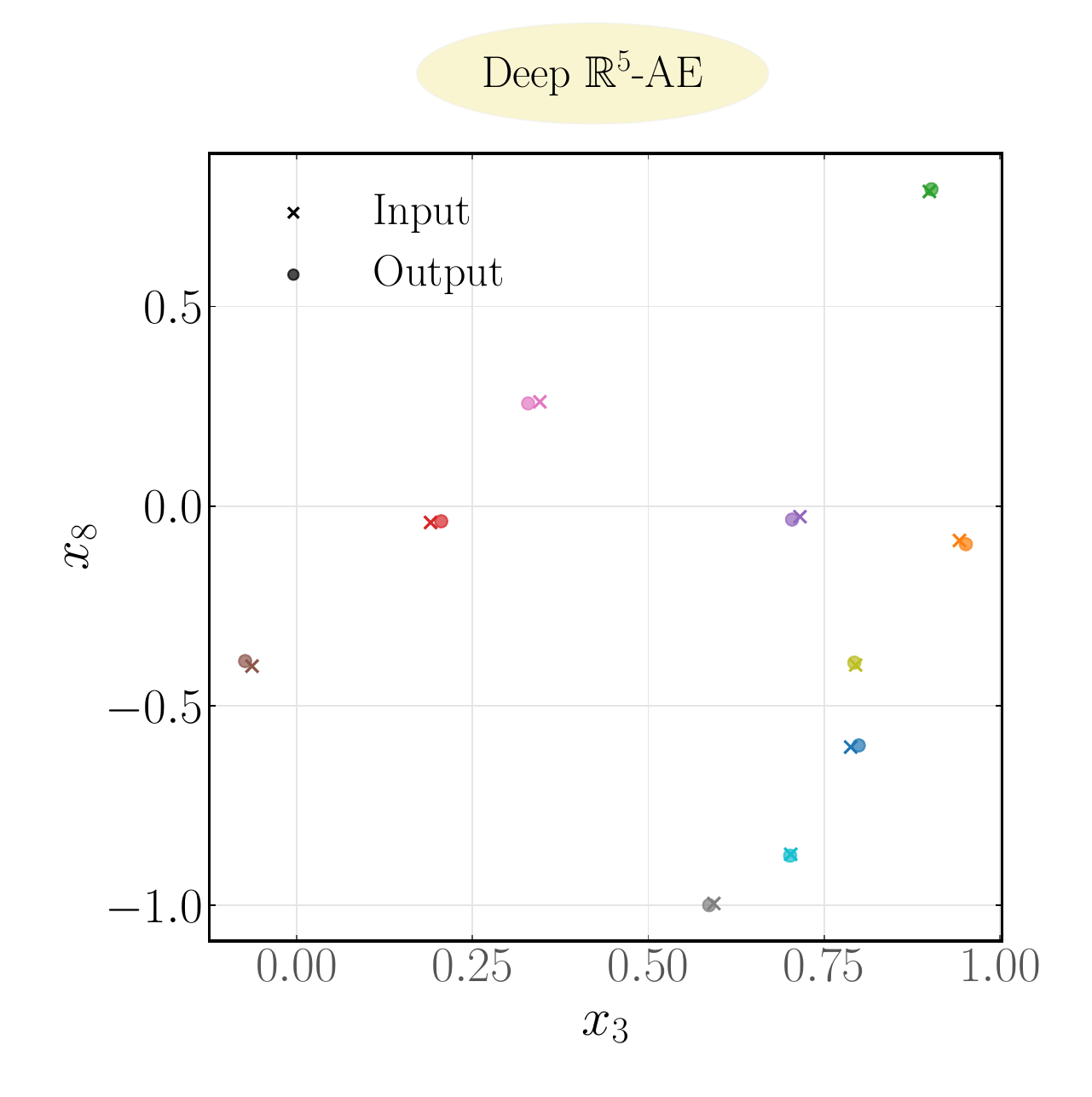}\\
     \includegraphics[scale=0.2]{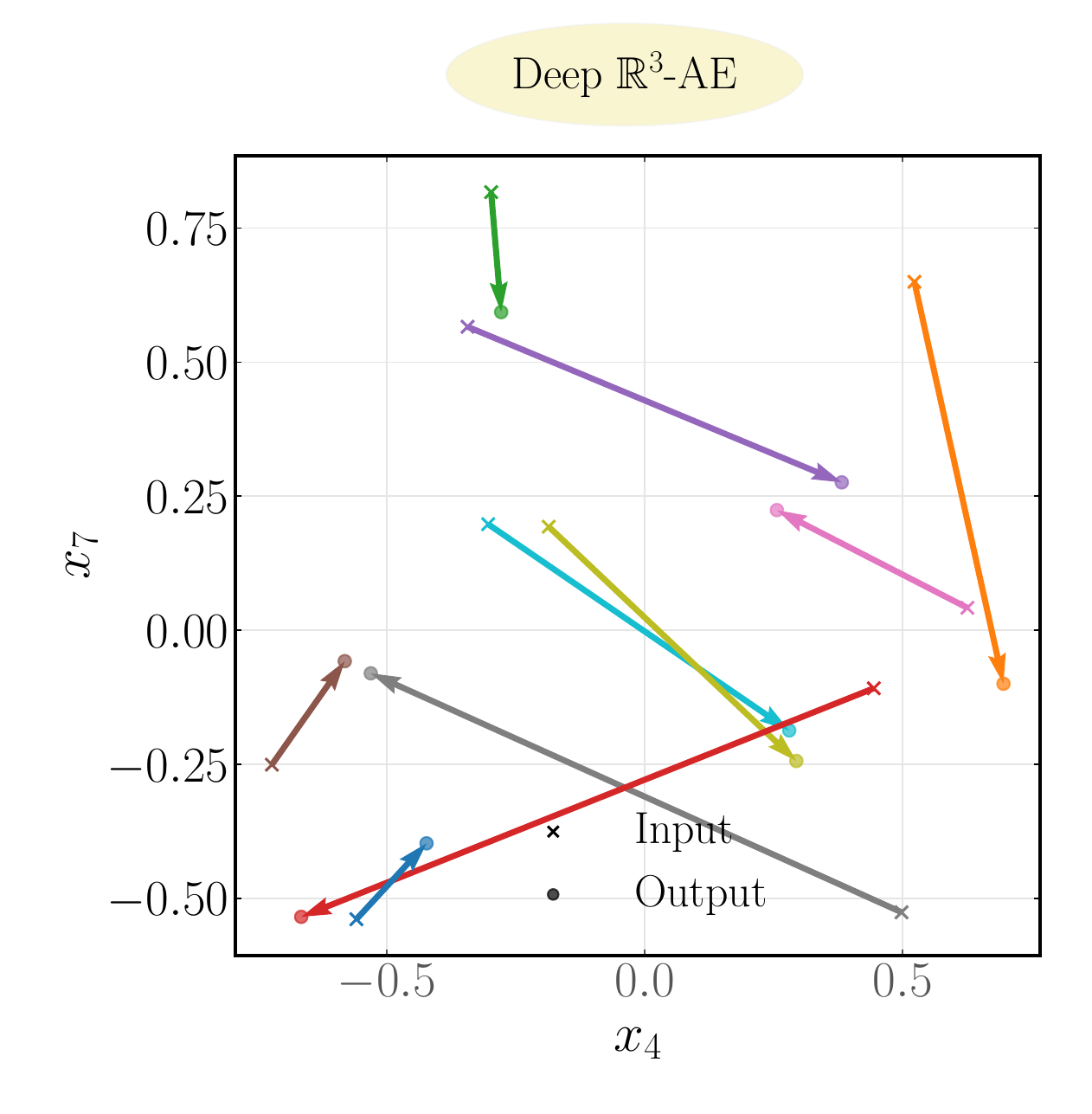}
    \includegraphics[scale=0.2]{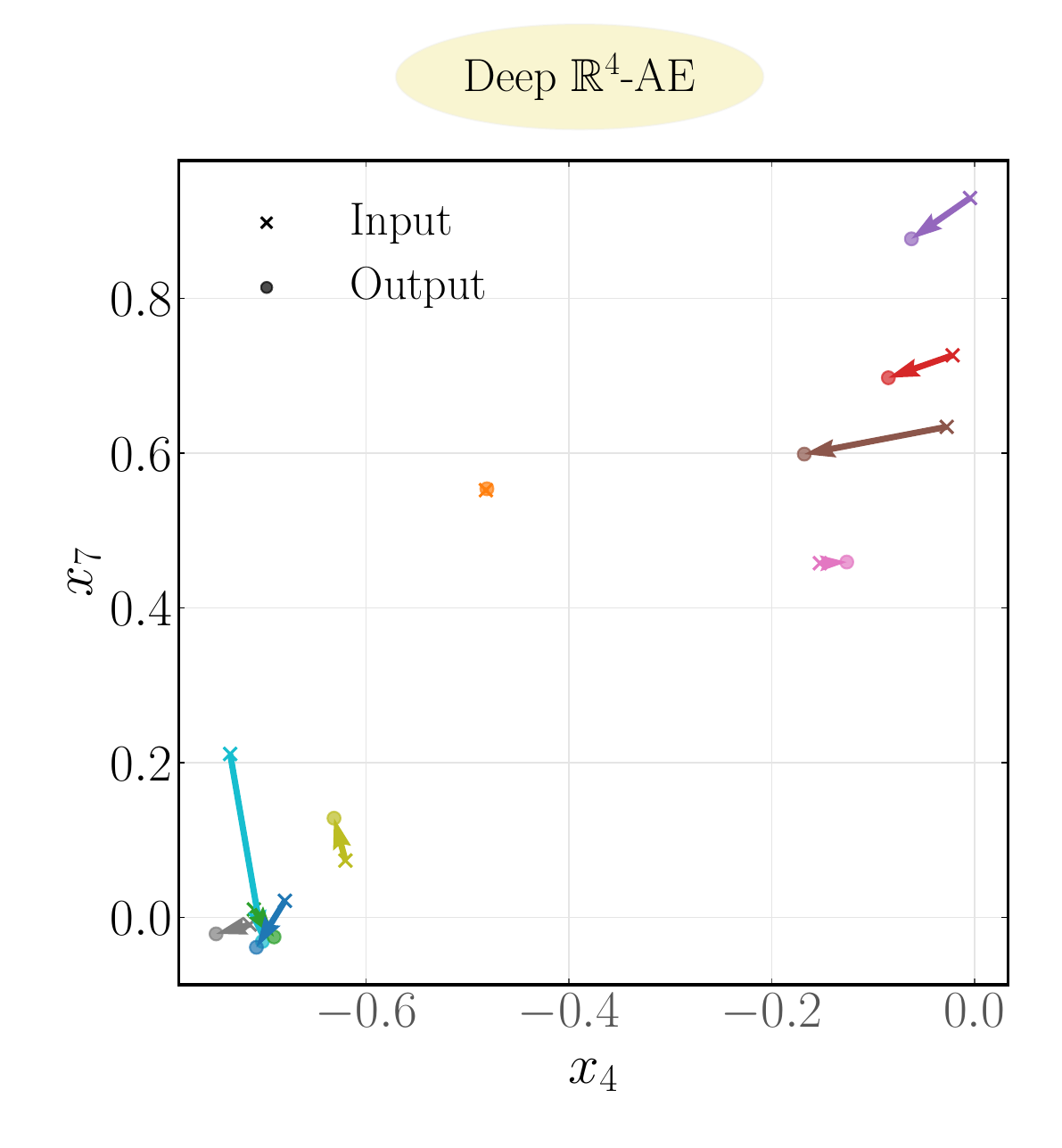}
    \includegraphics[scale=0.2]{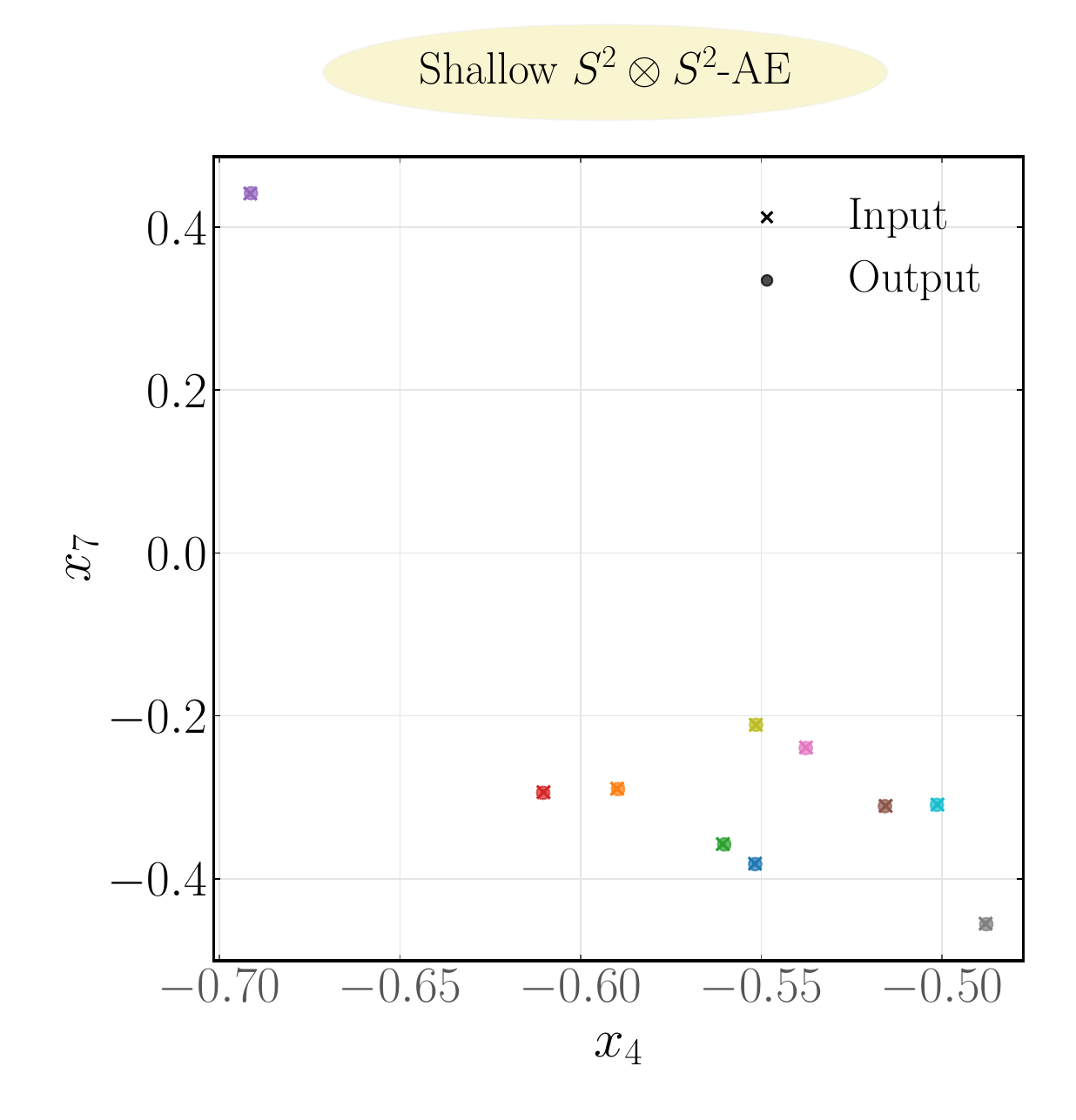}
    \includegraphics[scale=0.2]{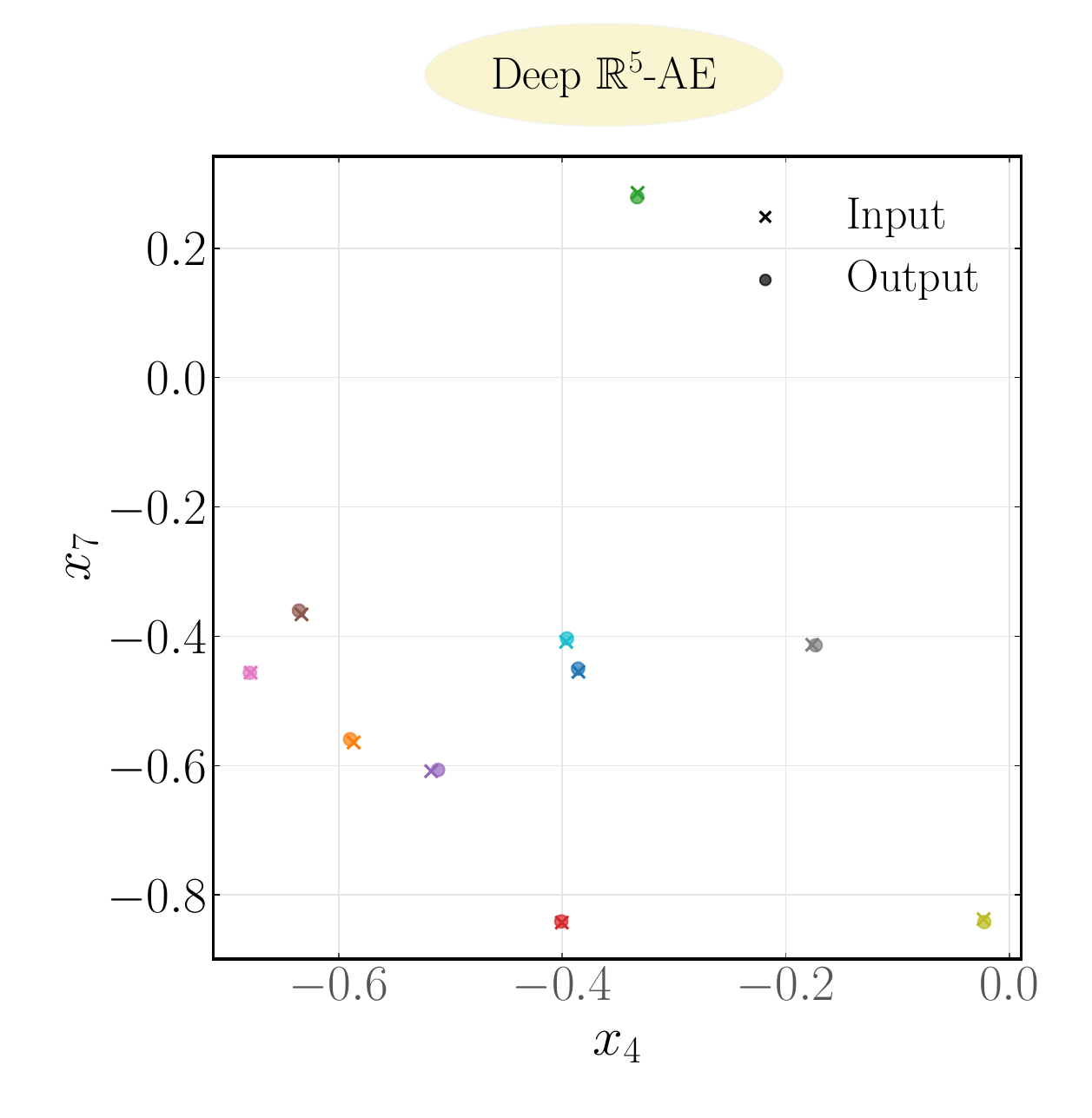}
    
	\caption{Input - output pairs for ten samples with highest reconstruction error for different latent spaces. The dataset is $S^2\otimes S^2$ embedded in $\mathbb{R}^9$. $9D$ data is projected orthogonally onto selected 2-axes combinations.}
	\label{fig:s2xs2_worst_point_scatter}
\end{figure}
}
 
\subsection{$S^2\otimes S^2$ embedded in $\mathbb{R}^9$  }
Before considering simulated collider data, we consider the case where the difference between the embedded space's extrinsic dimensions and the data manifold's intrinsic dimension is more than one.  Concretely, we take the product of two 2-spheres $S^2\otimes S^2$ embedded in $\mathbb{R}^9$. This is done by first generating two independent embeddings of a 2-sphere in $\mathbb{R}^3$ with coordinates denoted as  $\mathbf{r}_a \in {\mathbb R}^3_a$ and $\mathbf{r}_b \in {\mathbb R}^3_b$, respectively. 
One then constructs the vector $\mathbf{X}'=\mathbf{r}_a \oplus\mathbf{r}_b \oplus \mathbf{0}_c$, with $\mathbf{0}_c$ being the zero-vector in $\in {\mathbb R}^3_c$.  The final sample in $\mathbb{R}^9 = \mathbb{R}_a^3 \otimes \mathbb{R}_b^3 \otimes \mathbb{R}_c^3$ is then obtained by the rotation $$\mathbf{X}=\mathbf{R}_{67}(\theta_6)\mathbf{R}_ {58}(\theta_5) \mathbf{R}_{49}(\theta_4)\mathbf{R}_ {39}(\theta_3)\mathbf{R}_{28}(\theta_2)\mathbf{R}_ {17}(\theta_1)\mathbf{X}'\;,$$
where $\mathbf{R}_{ij}(\theta)$ denotes a $9\times 9$ orthogonal matrix with $R_{ii}=R_{jj}=\cos\theta$ and $R_{kk}=1$ for $k\notin \{i,j\}$ and $R_{ij}=\sin\theta$ and $R_{ji}=-\sin\theta$ and all other non-diagonal entries being zero.  The rotation angles are taken to be $\theta_1=2.6994$, $\theta_2=2.3480$, $\theta_3=3.0390$, $\theta_4=3.9448$, $\theta_5=4.3052$, and $\theta_6=0.8107$. With these values fixed for the whole generated dataset, the data manifold is $S^2\otimes S^2$ while the manifold is embedded in $\mathbb{R}^9$. This setup closely mimics our background example of sequential 2-body decay of the top quark at the parton level with zero widths. In the following, the components of $\mathbf{X}$ are denoted as $x_i$ for $i\in\{1,2,...,9\}$.

On 400k training and 100k validation data, we train four autoencoders: three $\mathbb{R}^k$-AEs for  $k=\{3,4,5\}$,  and $S^2\otimes S^2$-AE. They have the same architectures as the corresponding AEs in the $S^2$ experiment, with the only essential difference coming in the input, latent and output layers. 
The $S^2\otimes S^2$ latent layer has extrinsic dimensions of six where the two triplets are independently normalised (as explained in section \ref{sec:top_embed}), reducing the number of degrees of freedom into four. 
For each of these autoencoders, a 2D projected scatter plot of the input and the output of ten samples with the highest reconstruction error within the validation data are shown in Fig.~\ref{fig:s2xs2_worst_point_scatter} in the $(x_2,x_3),\,(x_3,x_8)$ and $(x_4,x_7)$ planes. One notices that reconstruction is the worst in $\mathbb{R}^3$-AE despite the architecture having 1.4 million parameters since there does not exist any well-behaved local chart of $S^2\otimes S^2$ in $\mathbb{R}^3$. Since there are good local charts in $\mathbb{R}^4$, the situation is relatively better in $\mathbb{R}^4$-AE. 
However, points with large reconstruction errors are not completely removed since $\mathbb{R}^4$ cannot admit a global embedding of $S^2\otimes S^2$. For $S^2\otimes S^2$-AE, a small network of around 1k parameters can reconstruct the validation data efficiently with overlapping inputs and outputs even for the ten worst reconstructed samples in the validation dataset. Additionally, since $S^2\otimes S^2$ can be faithfully embedded in $\mathbb{R}^5$, we see that $\mathbb{R}^5$-AE can also efficiently reconstruct the validation data entirely.

 \section{Latent Topology-based Anomaly Detection}
 \label{sec:latentano}
 
 As seen above, manually inducing the desired topology in the latent space clears topological obstructions due to the non-trivial global properties of the data manifold while explicitly keeping the intrinsic dimensions consistent with the background data manifold.  In this section, we evaluate the anomaly detection performance of autoencoders with latent topological priors, taking an example of hadronic sequential two-body decay of the top quark as the background and the top quark's three-body decay via an effective four-point fermion operator to a bottom and two light quarks as a possible signal.

 \subsection{Event Simulation and Baseline Selection}
 For both background and signal, we consider top pair production where the anti-top decays sequentially to a bottom and $W^-$, with the $W^-$ decaying to a charged lepton and an antineutrino, while we concentrate on different decays of the top quark in the hadronic final state consisting of two light quarks and a bottom quark.  
 Parton level events for the background and signal are generated by \texttt{MadGraph5\_aMC@NLO (v3.5.5)}~\cite{Alwall:2014hca} at 14 TeV centre-of-mass energy. 
The ``reconstructed'' data samples are obtained from parton level ones by applying parton showering and hadronisation with \texttt{Pythia8 (v8.3.10)}\cite{Bierlich:2022pfr} and the event selection described below. 
 For the background, we generate semileptonic top decays of a pair of top quarks, where the $W^-$ decays leptonically, and the $W^+$ decays hadronically. For the signal, we use the \texttt{SMEFTsim\_top\_MwScheme\_UFO-massless} model from the \texttt{SMEFTsim}~\cite{Brivio:2017btx,Brivio:2020onw} package. Since we are only interested in three body decay of the top quark, the production of the top quark pair proceeds via QCD diagrams alone, and we force the top quark to decay without a resonant $W^+$ with the syntax \texttt{t > b j j NP=1 \$\$w+}, while the anti-top decays similarly to the background. During the showering and hadronisation stage, we keep all $B$ mesons stable after hadronisation by setting the user-flag \texttt{mayDecay = off} for simulating $b$-jet tagging. 

\begin{figure}
\centering 
\includegraphics[scale=0.2]{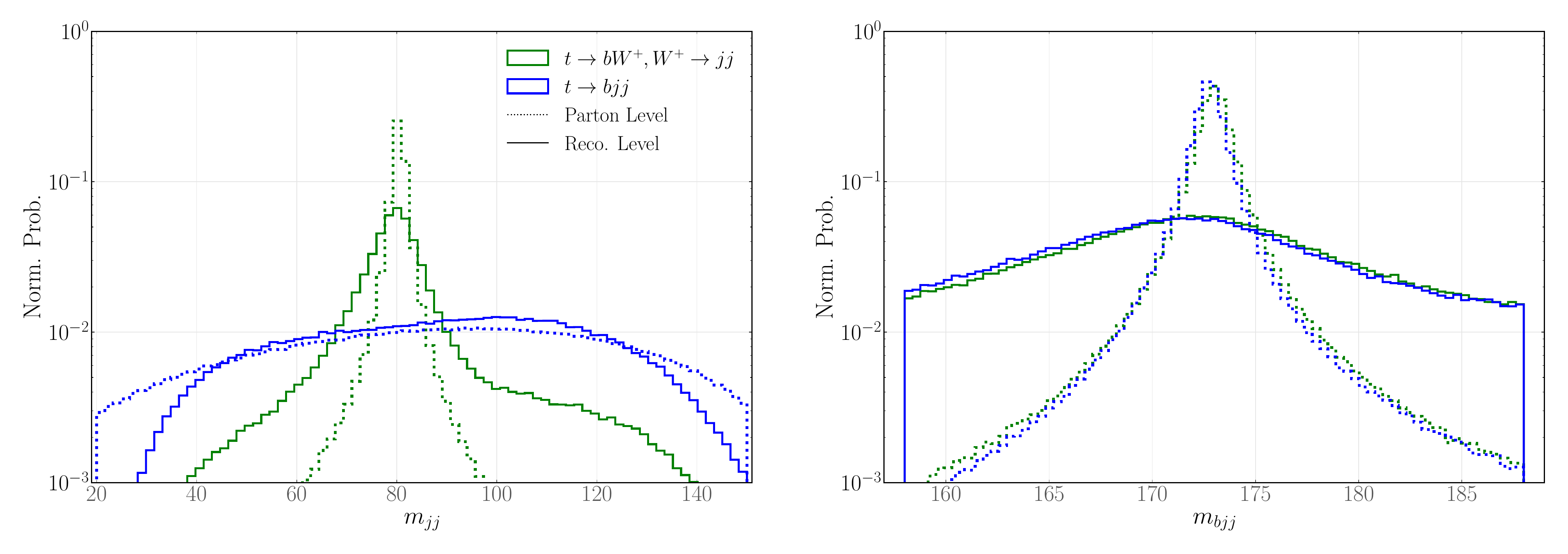} 
\caption{Invariant mass distribution of the $W^+$ $(m_{jj})$ and top $(m_{bjj})$ decay products after reconstruction and baseline selection (solid lines) and the true distribution at parton level (dotted lines). }
\label{fig:mass_plot} 
\end{figure}

 We select events via a basic event reconstruction and baseline selection utilising all stable final state particles obtained after hadronisation.  The event's missing transverse energy $\slashed{E}_T$ is evaluated as the transverse momentum magnitude of the four-vector sum of all particles after filtering out neutrinos and anti-neutrinos. We keep events with $\slashed{E}_T\geq 50$ GeV. Next, we reconstruct isolated leptons and photons by the isolation variable $I=(\sum_{\Delta R_{ia}<0.4} p_T^i)/p^a_T$, where $a$ is the particle having PID of either an electron, muon or a photon, and $i$ is any other particle within $\Delta R_{ia}<0.4$. If $I<0.12$, we consider particle $a$ to be isolated and remove all particles $i$ from further processing.  Events are selected if there is only one negatively charged lepton within pseudorapidity $|\eta|\leq 3$. Remaining final state particles are used to reconstruct jets of radius $R=0.4$ and  $p^{\text{min}}_T= 40$ GeV, with the anti-$k_t$ algorithm~\cite{Cacciari:2008gp} using \texttt{FastJet (v3.4.0)}~\cite{Cacciari:2011ma}. We select events with at least four jets within $|\eta| < 5$. We perform a simplistic $b$-jet tagging on all reconstructed jets by assigning a positive $b$-tag if there is a $B$-meson within $\Delta R <0.2$ of the jet axis. Events are selected if there are at least two $b$-tagged jets. After this, we ignore events if it has less than two untagged jets.  Using the two hardest untagged jets, we construct candidate top quarks with each $b$-tagged jet and ignore an event if none of their masses falls within 15 GeV of $m_{\text{top}}=173$ GeV. If more than one $b$-jet falls within the mass window, we select the one closest to $m_{\text{top}}$. 
The invariant mass of the two unflavoured jets and the selected top candidate are shown in Fig.~\ref{fig:mass_plot} along with the true distribution at parton level events.   

 We have 300k samples for the background divided into 240k training, 60k validation/testing samples, and 60k test samples for the signal.  To consider the effect of reconstruction and differences in the width of the initial particles, we also consider 500k parton level samples without any additional selection criteria, divided into 400k training and 100k validation/testing samples for the background and 100k testing samples for the signal.  For both these cases, we extract three-momentum of the three decay products in the top candidate's rest frame as input features embedded in $\mathbb{R}^9$, where the untagged jets are ordered by transverse momentum in the lab frame for reconstructed data. Each feature in the input is $z$-score standardised using the \texttt{StandardScaler} in \texttt{scikit-learn}~\cite{scikit-learn} package based on the mean and standard deviation of the entire background datasets.

\subsection{Network Architecture and Training}
\label{sec:net_arch_train}
Assuming distinguishable decay products and zero width decays in the the top's rest frame, the background data manifold is $S^2\otimes S^2$ while the signal's is $S^5$. Note that in this idealised limit, the intrinsic dimensions of the background manifold are four, while that of the signal manifold is five.   
Therefore, to consider the interplay of trivial latent topologies and the existence of local charts and global embeddings, we consider three $\mathbb{R}^k$-AEs for $k\in\{3,4,5\}$. Since the $W^+$-boson's decay manifests as two jets, which are experimentally indistinguishable, we note that the background manifold is $S^2\otimes \mathds{RP}^2$. Therefore, we consider three non-trivial latent topologies: $S^2\otimes S^2$, $S^2\otimes \mathds{RP}^2$, and $S^5$. 

The network analyses are carried out using \texttt{PyTorch (v2.0.0)}~\cite{10.5555/3454287.3455008} on a single NVIDIA A100 GPU.
All autoencoders have the same architecture as the $\mathbb{R}^k$-AEs considered in the experiments on toy datasets (see section~\ref{sec:toy_clear}) with 1.4 million tunable parameters. For architectures with non-trivial latent topologies, the extrinsic dimensions of the embeddings described in section~\ref{sec:top_embed} determine the number of nodes in the output and input of the encoder and decoder, respectively. For those with product topologies, the extrinsic dimensions are the sum of the two constituents, as the embeddings are constructed on mutually exclusive node outputs. We train each network on the background dataset using the mean squared error loss function ten times from random initialisation for a maximum of three hundred epochs via the \texttt{Adam}~\cite{DBLP:journals/corr/KingmaB14} optimiser with an initial learning rate of 0.001 on batches of 256 samples. The learning rate is reduced by half if the validation loss has not been reduced for three epochs, and the training stops if the loss has not improved for twenty epochs.

\subsection{Results} 
From each training instance, we choose the epoch with minimum validation loss and extract the reconstruction error on the test datasets of both background and signal classes. For each base architecture, the mean and standard deviation of the minimum validation loss, and area under the receiver-operator-characteristics curve (AUC) is shown in table~\ref{tab:parton_auc} and~\ref{tab:reco_auc} for the parton data and reconstructed data, respectively. The behaviour of the different topologies is similar for both datasets, with the difference in anomaly detection coming from the parton shower inducing a much larger width of the $W^+$ mass peak for the background samples in the reconstructed level data. From the optimisation perspective, heuristically, the best validation loss indicates the quality of the minima. 
The best reconstruction error is found when the latent dimensions are larger than four, i.e.\ $\mathbb{R}^5$-AE and $S^5$-AE, with at least an order-of-magnitude reduction from other topologies. 
However, this does not translate to better anomaly detection performance, as evidenced by their AUCs. 
As explained previously and considering the ideal manifold structure of the background and signal, the five-dimensional bottleneck layer will be able to represent local charts of the five-dimensional signal manifold even though it may not admit a global embedding,
which in turn leads to worse anomaly detection performance, as it can locally reconstruct the signal data well. 
Even though the latent space in $S^5$-AE is homeomorphic to the signal manifold, after training with the background data, it does not lead to an efficient reconstruction of the signal data. 
We see from the $W^+$ invariant mass distribution that the overlap between the background and signal is much larger for the reconstructed data. 
As a consequence, the AUC value for the reconstructed data in $S^5$-AE is smaller than the parton level one, as we can see in the tables.
Similarly, $\mathbb{R}^3$, which does not admit local charts of the four-dimensional background manifold, has a higher reconstruction error than all other latent topologies. 
Therefore, its anomaly detection capabilities do not quite match those of four-dimensional ones while being better than five dimensional ones on the reconstructed data and nominally poorer for the parton-level data.

All four-dimensional latent topologies have the best validation losses compared to the five-dimensional and three-dimensional ones. Within them, the lowest and second lowest validation loss are for $S^2\otimes S^2$ and $S^2\otimes \mathds{RP}^2$ latent topologies, respectively. Clearly, this shows that the simplified assumptions of the background data manifold hold true to a good extent. While we expect the background data manifold to be closest to $S^2\otimes \mathds{RP}^2$, there are two possible reasons for its slightly lower reconstruction efficiency compared to $S^2\otimes S^2$. Firstly, our network does not explicitly respect the discrete symmetry of the identification $(\mathbf{p}_a,\mathbf{p}_b)\sim(\mathbf{p}_b,\mathbf{p}_a)$ since the network output is not invariant under the exchange of the two momenta.\footnote{However, we expect the dataset to contain equal amounts of $(\mathbf{p}_a,\mathbf{p}_b)$ and $(\mathbf{p}_b,\mathbf{p}_a)$ and therefore approximately respect the symmetry for large enough data. 
} Secondly, $\mathds{RP}^2$ has a relatively more complicated structure than $S^2$ which could result in a harder optimisation task. The anomaly detection performance of the $S^2\otimes \mathds{RP}^2$ is nominally higher for the reconstructed data while being identical within the specified precision for the parton-level data. 
For the $\mathbb{R}^4$-AE, we see that it has the worst validation within four-dimensional topologies. 
 with the non-trivial four-dimensional topologies having comparatively better anomaly detection performance for either datasets. This confirms that designing and using a non-trivial latent topology homeomorphic to the background data manifold can help improve anomaly detection performance.   

\begin{table} 

\begin{tabular}{lll}
\toprule
Latent Topology & Best Val. loss & AUC \\
\midrule
$\mathbb{R}^3$ & 0.0329$\pm$0.0045 & 0.8449$\pm$0.0082 \\
$\mathbb{R}^4$ & 0.0026$\pm$0.0008 & 0.9417$\pm$0.0054 \\
$\mathbb{R}^5$ & 0.0003$\pm$0.0000  & 0.8700$\pm$0.0035 \\
$S^{2}\otimes S^2$ & 0.0019$\pm$0.0000  & 0.9466$\pm$0.0002 \\
$S^2\otimes\mathds{RP}^2$ & 0.0019$\pm$0.0000 & 0.9466$\pm$0.0002 \\
$S^5$ & 0.0003$\pm$0.0000 & 0.8612$\pm$0.0063 \\
\bottomrule
\end{tabular}

\caption{Mean and standard deviation over ten training instances of the best validation loss, and AUC on the signal dataset for autoencoders with different latent topologies trained and tested on the parton-level data.}
\label{tab:parton_auc}
 \end{table}

\begin{table}
%\resizebox{0.8\textwidth}{!}{
\begin{tabular}{lll}
\toprule
Latent Topology & Best Val. loss & AUC \\
\midrule
$\mathbb{R}^3$ & 0.0589$\pm$0.0022  & 0.6939$\pm$0.0048 \\
$\mathbb{R}^4$ & 0.0172$\pm$0.0026  & 0.7246$\pm$0.0097 \\
$\mathbb{R}^5$ & 0.0019$\pm$0.0000 & 0.5490$\pm$0.0015 \\
$S^{2}\otimes S^2$ & 0.0136$\pm$0.0009  & 0.7316$\pm$0.0047 \\
$S^2\otimes\mathds{RP}^2$ & 0.0145$\pm$0.0014 & 0.7321$\pm$0.0037 \\
$S^5$ & 0.0019$\pm$0.0000 & 0.5450$\pm$0.0011 \\
\bottomrule
\end{tabular}

%}

\caption{Mean and standard deviation over ten training instances of the best validation loss, and AUC on the signal dataset for autoencoders with different latent topologies trained and tested on the reconstructed data with baseline selection cuts.}
\label{tab:reco_auc}

\end{table}

%===================================================================
\section{Conclusions}
\label{sec:conclusions}
%===================================================================
In this work, we have introduced and systematically studied topology-aware autoencoders for anomaly detection in high-energy physics datasets. Our approach addresses the fundamental issue that arises when autoencoder latent spaces fail to capture the non-trivial topology of phase-space manifolds in particle scattering and decay processes. By constructing autoencoders with latent spaces that explicitly respect the topological structure of the data, we demonstrated that anomaly detection can be significantly improved.

We established a theoretical framework for embedding non-trivial background manifolds into latent spaces with matching topology. Our study covered several physical scenarios, including manifolds of type $S^n$, $S^n \otimes S^m$ and $\mathds{RP}^2$, which naturally emerge in multi-body decays and scattering processes. Through numerical experiments, we verified that topology-aware latent spaces mitigate topological obstructions present in conventional autoencoders, which often result in spurious high reconstruction errors for physically allowed background regions. 

Our numerical results confirm the effectiveness of our approach. Using synthetic datasets, we first demonstrated that when the $S^2$ data is applied to the ${\mathbb R}^2$-AE,  
the reconstruction error exhibits distinct peaks in the loss-versus-distance plot, indicating a topological mismatch. 
In contrast, topology-aware latent spaces reduce reconstruction errors and allow for a more faithful representation of the background manifold. 
For example, in the case of an $S^2 \otimes S^2$ data presented in ${\mathbb R}^9$, we found that conventional autoencoders with ${\mathbb R}^4$ latent spaces struggled to accurately reconstruct the data, enhancing the possibility of artificial classification of background events as anomalous. 
By explicitly constructing an $S^2 \otimes S^2$ latent topology, we achieved near-perfect reconstruction of the background data, demonstrating the advantage of incorporating topological priors.

Applying our method to simulated collider data, we found that autoencoders with topology-aware latent manifolds outperformed traditional architectures in anomaly detection. Specifically, for the hadronic decay of the top quark, where the background phase space is well approximated by $S^2 \otimes \mathds{RP}^2$, our proposed architecture achieved an AUC of $0.7321 \pm 0.0037$, compared to $0.7246 \pm 0.0097$ for an autoencoder with a trivial $\mathbb{R}^4$ latent space. We also found that higher-dimensional trivial latent spaces, such as $\mathbb{R}^5$, led to significantly lower AUC values ($0.549 \pm 0.0015$), indicating that unconstrained latent dimensions allow for efficient reconstruction of both background and signal, thereby degrading anomaly detection performance. These results underscore the importance of carefully selecting the latent space topology to maximize sensitivity to out-of-distribution events.

Beyond improvements in anomaly detection, our findings highlight a fundamental connection between data topology and deep learning representations. Our approach provides a systematic framework to incorporate physically relevant priors in machine learning architectures, paving the way for more robust and interpretable anomaly detection methods in experimental particle physics. 

Future work could explore dynamic latent space topology adaptation to handle scenarios where the background topology is not known a priori. 
Additionally, our framework can be extended to study more complex event topologies and multi-dimensional manifolds in experimental datasets. 
Given the promising performance of topology-aware anomaly detection, we anticipate its broader adoption in searches for new physics at colliders and other high-dimensional scientific datasets.

%===================================================================
\section*{Acknowledgements}
VSN and MS are supported by STFC under grant ST/X003167/1.
%===================================================================

%%%%%%%%%%%%%%%%%%%%%%%%%%%%%%%%%%%%%%%%%%%%%%%%%%

%%%%%%%%%%%%%%%%%%%%%%%%%%%%%%%%%%%%%%%%%%%%%%%%%%
\bibliography{refs}

\begin{thebibliography}{74}
\expandafter\ifx\csname natexlab\endcsname\relax\def\natexlab#1{#1}\fi
\expandafter\ifx\csname bibnamefont\endcsname\relax
  \def\bibnamefont#1{#1}\fi
\expandafter\ifx\csname bibfnamefont\endcsname\relax
  \def\bibfnamefont#1{#1}\fi
\expandafter\ifx\csname citenamefont\endcsname\relax
  \def\citenamefont#1{#1}\fi
\expandafter\ifx\csname url\endcsname\relax
  \def\url#1{\texttt{#1}}\fi
\expandafter\ifx\csname urlprefix\endcsname\relax\def\urlprefix{URL }\fi
\providecommand{\bibinfo}[2]{#2}
\providecommand{\eprint}[2][]{\url{#2}}

\bibitem[{\citenamefont{Belis et~al.}(2024)\citenamefont{Belis, Odagiu, and
  Aarrestad}}]{Belis:2023mqs}
\bibinfo{author}{\bibfnamefont{V.}~\bibnamefont{Belis}},
  \bibinfo{author}{\bibfnamefont{P.}~\bibnamefont{Odagiu}}, \bibnamefont{and}
  \bibinfo{author}{\bibfnamefont{T.~K.} \bibnamefont{Aarrestad}},
  \bibinfo{journal}{Rev. Phys.} \textbf{\bibinfo{volume}{12}},
  \bibinfo{pages}{100091} (\bibinfo{year}{2024}), \eprint{2312.14190}.

\bibitem[{\citenamefont{Kasieczka et~al.}(2021)}]{Kasieczka:2021xcg}
\bibinfo{author}{\bibfnamefont{G.}~\bibnamefont{Kasieczka}}
  \bibnamefont{et~al.}, \bibinfo{journal}{Rept. Prog. Phys.}
  \textbf{\bibinfo{volume}{84}}, \bibinfo{pages}{124201}
  (\bibinfo{year}{2021}), \eprint{2101.08320}.

\bibitem[{\citenamefont{Aarrestad et~al.}(2022)}]{Aarrestad:2021oeb}
\bibinfo{author}{\bibfnamefont{T.}~\bibnamefont{Aarrestad}}
  \bibnamefont{et~al.}, \bibinfo{journal}{SciPost Phys.}
  \textbf{\bibinfo{volume}{12}}, \bibinfo{pages}{043} (\bibinfo{year}{2022}),
  \eprint{2105.14027}.

\bibitem[{\citenamefont{D'Agnolo and Wulzer}(2019)}]{DAgnolo:2018cun}
\bibinfo{author}{\bibfnamefont{R.~T.} \bibnamefont{D'Agnolo}} \bibnamefont{and}
  \bibinfo{author}{\bibfnamefont{A.}~\bibnamefont{Wulzer}},
  \bibinfo{journal}{Phys. Rev. D} \textbf{\bibinfo{volume}{99}},
  \bibinfo{pages}{015014} (\bibinfo{year}{2019}), \eprint{1806.02350}.

\bibitem[{\citenamefont{Collins et~al.}(2018)\citenamefont{Collins, Howe, and
  Nachman}}]{Collins:2018epr}
\bibinfo{author}{\bibfnamefont{J.~H.} \bibnamefont{Collins}},
  \bibinfo{author}{\bibfnamefont{K.}~\bibnamefont{Howe}}, \bibnamefont{and}
  \bibinfo{author}{\bibfnamefont{B.}~\bibnamefont{Nachman}},
  \bibinfo{journal}{Phys. Rev. Lett.} \textbf{\bibinfo{volume}{121}},
  \bibinfo{pages}{241803} (\bibinfo{year}{2018}), \eprint{1805.02664}.

\bibitem[{\citenamefont{Collins et~al.}(2019)\citenamefont{Collins, Howe, and
  Nachman}}]{Collins:2019jip}
\bibinfo{author}{\bibfnamefont{J.~H.} \bibnamefont{Collins}},
  \bibinfo{author}{\bibfnamefont{K.}~\bibnamefont{Howe}}, \bibnamefont{and}
  \bibinfo{author}{\bibfnamefont{B.}~\bibnamefont{Nachman}},
  \bibinfo{journal}{Phys. Rev. D} \textbf{\bibinfo{volume}{99}},
  \bibinfo{pages}{014038} (\bibinfo{year}{2019}), \eprint{1902.02634}.

\bibitem[{\citenamefont{Hajer et~al.}(2020)\citenamefont{Hajer, Li, Liu, and
  Wang}}]{Hajer:2018kqm}
\bibinfo{author}{\bibfnamefont{J.}~\bibnamefont{Hajer}},
  \bibinfo{author}{\bibfnamefont{Y.-Y.} \bibnamefont{Li}},
  \bibinfo{author}{\bibfnamefont{T.}~\bibnamefont{Liu}}, \bibnamefont{and}
  \bibinfo{author}{\bibfnamefont{H.}~\bibnamefont{Wang}},
  \bibinfo{journal}{Phys. Rev. D} \textbf{\bibinfo{volume}{101}},
  \bibinfo{pages}{076015} (\bibinfo{year}{2020}), \eprint{1807.10261}.

\bibitem[{\citenamefont{De~Simone and Jacques}(2019)}]{DeSimone:2018efk}
\bibinfo{author}{\bibfnamefont{A.}~\bibnamefont{De~Simone}} \bibnamefont{and}
  \bibinfo{author}{\bibfnamefont{T.}~\bibnamefont{Jacques}},
  \bibinfo{journal}{Eur. Phys. J. C} \textbf{\bibinfo{volume}{79}},
  \bibinfo{pages}{289} (\bibinfo{year}{2019}), \eprint{1807.06038}.

\bibitem[{\citenamefont{Andreassen et~al.}(2020)\citenamefont{Andreassen,
  Nachman, and Shih}}]{Andreassen:2020nkr}
\bibinfo{author}{\bibfnamefont{A.}~\bibnamefont{Andreassen}},
  \bibinfo{author}{\bibfnamefont{B.}~\bibnamefont{Nachman}}, \bibnamefont{and}
  \bibinfo{author}{\bibfnamefont{D.}~\bibnamefont{Shih}},
  \bibinfo{journal}{Phys. Rev. D} \textbf{\bibinfo{volume}{101}},
  \bibinfo{pages}{095004} (\bibinfo{year}{2020}), \eprint{2001.05001}.

\bibitem[{\citenamefont{Nachman and Shih}(2020)}]{Nachman:2020lpy}
\bibinfo{author}{\bibfnamefont{B.}~\bibnamefont{Nachman}} \bibnamefont{and}
  \bibinfo{author}{\bibfnamefont{D.}~\bibnamefont{Shih}},
  \bibinfo{journal}{Phys. Rev. D} \textbf{\bibinfo{volume}{101}},
  \bibinfo{pages}{075042} (\bibinfo{year}{2020}), \eprint{2001.04990}.

\bibitem[{\citenamefont{Knapp et~al.}(2021)\citenamefont{Knapp, Cerri,
  Dissertori, Nguyen, Pierini, and Vlimant}}]{Knapp:2020dde}
\bibinfo{author}{\bibfnamefont{O.}~\bibnamefont{Knapp}},
  \bibinfo{author}{\bibfnamefont{O.}~\bibnamefont{Cerri}},
  \bibinfo{author}{\bibfnamefont{G.}~\bibnamefont{Dissertori}},
  \bibinfo{author}{\bibfnamefont{T.~Q.} \bibnamefont{Nguyen}},
  \bibinfo{author}{\bibfnamefont{M.}~\bibnamefont{Pierini}}, \bibnamefont{and}
  \bibinfo{author}{\bibfnamefont{J.-R.} \bibnamefont{Vlimant}},
  \bibinfo{journal}{Eur. Phys. J. Plus} \textbf{\bibinfo{volume}{136}},
  \bibinfo{pages}{236} (\bibinfo{year}{2021}), \eprint{2005.01598}.

\bibitem[{\citenamefont{Aad et~al.}(2020)}]{ATLAS:2020iwa}
\bibinfo{author}{\bibfnamefont{G.}~\bibnamefont{Aad}} \bibnamefont{et~al.}
  (\bibinfo{collaboration}{ATLAS}), \bibinfo{journal}{Phys. Rev. Lett.}
  \textbf{\bibinfo{volume}{125}}, \bibinfo{pages}{131801}
  (\bibinfo{year}{2020}), \eprint{2005.02983}.

\bibitem[{\citenamefont{Dillon et~al.}(2020)\citenamefont{Dillon, Faroughy,
  Kamenik, and Szewc}}]{Dillon:2020quc}
\bibinfo{author}{\bibfnamefont{B.~M.} \bibnamefont{Dillon}},
  \bibinfo{author}{\bibfnamefont{D.~A.} \bibnamefont{Faroughy}},
  \bibinfo{author}{\bibfnamefont{J.~F.} \bibnamefont{Kamenik}},
  \bibnamefont{and} \bibinfo{author}{\bibfnamefont{M.}~\bibnamefont{Szewc}},
  \bibinfo{journal}{JHEP} \textbf{\bibinfo{volume}{10}}, \bibinfo{pages}{206}
  (\bibinfo{year}{2020}), \eprint{2005.12319}.

\bibitem[{\citenamefont{Crispim Rom\~ao et~al.}(2021)\citenamefont{Crispim
  Rom\~ao, Castro, and Pedro}}]{CrispimRomao:2020ucc}
\bibinfo{author}{\bibfnamefont{M.}~\bibnamefont{Crispim Rom\~ao}},
  \bibinfo{author}{\bibfnamefont{N.~F.} \bibnamefont{Castro}},
  \bibnamefont{and} \bibinfo{author}{\bibfnamefont{R.}~\bibnamefont{Pedro}},
  \bibinfo{journal}{Eur. Phys. J. C} \textbf{\bibinfo{volume}{81}},
  \bibinfo{pages}{27} (\bibinfo{year}{2021}), \bibinfo{note}{[Erratum:
  Eur.Phys.J.C 81, 1020 (2021)]}, \eprint{2006.05432}.

\bibitem[{\citenamefont{Cheng et~al.}(2023)\citenamefont{Cheng, Arguin,
  Leissner-Martin, Pilette, and Golling}}]{Cheng:2020dal}
\bibinfo{author}{\bibfnamefont{T.}~\bibnamefont{Cheng}},
  \bibinfo{author}{\bibfnamefont{J.-F.} \bibnamefont{Arguin}},
  \bibinfo{author}{\bibfnamefont{J.}~\bibnamefont{Leissner-Martin}},
  \bibinfo{author}{\bibfnamefont{J.}~\bibnamefont{Pilette}}, \bibnamefont{and}
  \bibinfo{author}{\bibfnamefont{T.}~\bibnamefont{Golling}},
  \bibinfo{journal}{Phys. Rev. D} \textbf{\bibinfo{volume}{107}},
  \bibinfo{pages}{016002} (\bibinfo{year}{2023}), \eprint{2007.01850}.

\bibitem[{\citenamefont{Khosa and Sanz}(2023)}]{Khosa:2020qrz}
\bibinfo{author}{\bibfnamefont{C.~K.} \bibnamefont{Khosa}} \bibnamefont{and}
  \bibinfo{author}{\bibfnamefont{V.}~\bibnamefont{Sanz}},
  \bibinfo{journal}{SciPost Phys.} \textbf{\bibinfo{volume}{15}},
  \bibinfo{pages}{053} (\bibinfo{year}{2023}), \eprint{2007.14462}.

\bibitem[{\citenamefont{Mikuni and Canelli}(2021)}]{Mikuni:2020qds}
\bibinfo{author}{\bibfnamefont{V.}~\bibnamefont{Mikuni}} \bibnamefont{and}
  \bibinfo{author}{\bibfnamefont{F.}~\bibnamefont{Canelli}},
  \bibinfo{journal}{Phys. Rev. D} \textbf{\bibinfo{volume}{103}},
  \bibinfo{pages}{092007} (\bibinfo{year}{2021}), \eprint{2010.07106}.

\bibitem[{\citenamefont{Park et~al.}(2020)\citenamefont{Park, Rankin, Udrescu,
  Yunus, and Harris}}]{Park:2020pak}
\bibinfo{author}{\bibfnamefont{S.~E.} \bibnamefont{Park}},
  \bibinfo{author}{\bibfnamefont{D.}~\bibnamefont{Rankin}},
  \bibinfo{author}{\bibfnamefont{S.-M.} \bibnamefont{Udrescu}},
  \bibinfo{author}{\bibfnamefont{M.}~\bibnamefont{Yunus}}, \bibnamefont{and}
  \bibinfo{author}{\bibfnamefont{P.}~\bibnamefont{Harris}},
  \bibinfo{journal}{JHEP} \textbf{\bibinfo{volume}{21}}, \bibinfo{pages}{030}
  (\bibinfo{year}{2020}), \eprint{2011.03550}.

\bibitem[{\citenamefont{Blance and Spannowsky}(2020)}]{Blance:2020ktp}
\bibinfo{author}{\bibfnamefont{A.}~\bibnamefont{Blance}} \bibnamefont{and}
  \bibinfo{author}{\bibfnamefont{M.}~\bibnamefont{Spannowsky}},
  \bibinfo{journal}{JHEP} \textbf{\bibinfo{volume}{21}}, \bibinfo{pages}{170}
  (\bibinfo{year}{2020}), \eprint{2103.03897}.

\bibitem[{\citenamefont{Dorigo et~al.}(2023)\citenamefont{Dorigo, Fumanelli,
  Maccani, Mojsovska, Strong, and Scarpa}}]{Dorigo:2021iyy}
\bibinfo{author}{\bibfnamefont{T.}~\bibnamefont{Dorigo}},
  \bibinfo{author}{\bibfnamefont{M.}~\bibnamefont{Fumanelli}},
  \bibinfo{author}{\bibfnamefont{C.}~\bibnamefont{Maccani}},
  \bibinfo{author}{\bibfnamefont{M.}~\bibnamefont{Mojsovska}},
  \bibinfo{author}{\bibfnamefont{G.~C.} \bibnamefont{Strong}},
  \bibnamefont{and} \bibinfo{author}{\bibfnamefont{B.}~\bibnamefont{Scarpa}},
  \bibinfo{journal}{JHEP} \textbf{\bibinfo{volume}{01}}, \bibinfo{pages}{008}
  (\bibinfo{year}{2023}), \eprint{2106.05747}.

\bibitem[{\citenamefont{Caron et~al.}(2022)\citenamefont{Caron, Hendriks, and
  Verheyen}}]{Caron:2021wmq}
\bibinfo{author}{\bibfnamefont{S.}~\bibnamefont{Caron}},
  \bibinfo{author}{\bibfnamefont{L.}~\bibnamefont{Hendriks}}, \bibnamefont{and}
  \bibinfo{author}{\bibfnamefont{R.}~\bibnamefont{Verheyen}},
  \bibinfo{journal}{SciPost Phys.} \textbf{\bibinfo{volume}{12}},
  \bibinfo{pages}{077} (\bibinfo{year}{2022}), \eprint{2106.10164}.

\bibitem[{\citenamefont{Hallin et~al.}(2022)\citenamefont{Hallin, Isaacson,
  Kasieczka, Krause, Nachman, Quadfasel, Schlaffer, Shih, and
  Sommerhalder}}]{Hallin:2021wme}
\bibinfo{author}{\bibfnamefont{A.}~\bibnamefont{Hallin}},
  \bibinfo{author}{\bibfnamefont{J.}~\bibnamefont{Isaacson}},
  \bibinfo{author}{\bibfnamefont{G.}~\bibnamefont{Kasieczka}},
  \bibinfo{author}{\bibfnamefont{C.}~\bibnamefont{Krause}},
  \bibinfo{author}{\bibfnamefont{B.}~\bibnamefont{Nachman}},
  \bibinfo{author}{\bibfnamefont{T.}~\bibnamefont{Quadfasel}},
  \bibinfo{author}{\bibfnamefont{M.}~\bibnamefont{Schlaffer}},
  \bibinfo{author}{\bibfnamefont{D.}~\bibnamefont{Shih}}, \bibnamefont{and}
  \bibinfo{author}{\bibfnamefont{M.}~\bibnamefont{Sommerhalder}},
  \bibinfo{journal}{Phys. Rev. D} \textbf{\bibinfo{volume}{106}},
  \bibinfo{pages}{055006} (\bibinfo{year}{2022}), \eprint{2109.00546}.

\bibitem[{\citenamefont{Mikuni et~al.}(2022)\citenamefont{Mikuni, Nachman, and
  Shih}}]{Mikuni:2021nwn}
\bibinfo{author}{\bibfnamefont{V.}~\bibnamefont{Mikuni}},
  \bibinfo{author}{\bibfnamefont{B.}~\bibnamefont{Nachman}}, \bibnamefont{and}
  \bibinfo{author}{\bibfnamefont{D.}~\bibnamefont{Shih}},
  \bibinfo{journal}{Phys. Rev. D} \textbf{\bibinfo{volume}{105}},
  \bibinfo{pages}{055006} (\bibinfo{year}{2022}), \eprint{2111.06417}.

\bibitem[{\citenamefont{d'Agnolo et~al.}(2022)\citenamefont{d'Agnolo, Grosso,
  Pierini, Wulzer, and Zanetti}}]{dAgnolo:2021aun}
\bibinfo{author}{\bibfnamefont{R.~T.} \bibnamefont{d'Agnolo}},
  \bibinfo{author}{\bibfnamefont{G.}~\bibnamefont{Grosso}},
  \bibinfo{author}{\bibfnamefont{M.}~\bibnamefont{Pierini}},
  \bibinfo{author}{\bibfnamefont{A.}~\bibnamefont{Wulzer}}, \bibnamefont{and}
  \bibinfo{author}{\bibfnamefont{M.}~\bibnamefont{Zanetti}},
  \bibinfo{journal}{Eur. Phys. J. C} \textbf{\bibinfo{volume}{82}},
  \bibinfo{pages}{275} (\bibinfo{year}{2022}), \eprint{2111.13633}.

\bibitem[{\citenamefont{Park et~al.}(2023)\citenamefont{Park, Harris, and
  Ostdiek}}]{Park:2022zov}
\bibinfo{author}{\bibfnamefont{S.~E.} \bibnamefont{Park}},
  \bibinfo{author}{\bibfnamefont{P.}~\bibnamefont{Harris}}, \bibnamefont{and}
  \bibinfo{author}{\bibfnamefont{B.}~\bibnamefont{Ostdiek}},
  \bibinfo{journal}{JHEP} \textbf{\bibinfo{volume}{07}}, \bibinfo{pages}{108}
  (\bibinfo{year}{2023}), \eprint{2208.05484}.

\bibitem[{\citenamefont{Hallin et~al.}(2023)\citenamefont{Hallin, Kasieczka,
  Quadfasel, Shih, and Sommerhalder}}]{Hallin:2022eoq}
\bibinfo{author}{\bibfnamefont{A.}~\bibnamefont{Hallin}},
  \bibinfo{author}{\bibfnamefont{G.}~\bibnamefont{Kasieczka}},
  \bibinfo{author}{\bibfnamefont{T.}~\bibnamefont{Quadfasel}},
  \bibinfo{author}{\bibfnamefont{D.}~\bibnamefont{Shih}}, \bibnamefont{and}
  \bibinfo{author}{\bibfnamefont{M.}~\bibnamefont{Sommerhalder}},
  \bibinfo{journal}{Phys. Rev. D} \textbf{\bibinfo{volume}{107}},
  \bibinfo{pages}{114012} (\bibinfo{year}{2023}), \eprint{2210.14924}.

\bibitem[{\citenamefont{Kasieczka et~al.}(2023)\citenamefont{Kasieczka,
  Mastandrea, Mikuni, Nachman, Pettee, and Shih}}]{Kasieczka:2022naq}
\bibinfo{author}{\bibfnamefont{G.}~\bibnamefont{Kasieczka}},
  \bibinfo{author}{\bibfnamefont{R.}~\bibnamefont{Mastandrea}},
  \bibinfo{author}{\bibfnamefont{V.}~\bibnamefont{Mikuni}},
  \bibinfo{author}{\bibfnamefont{B.}~\bibnamefont{Nachman}},
  \bibinfo{author}{\bibfnamefont{M.}~\bibnamefont{Pettee}}, \bibnamefont{and}
  \bibinfo{author}{\bibfnamefont{D.}~\bibnamefont{Shih}},
  \bibinfo{journal}{Phys. Rev. D} \textbf{\bibinfo{volume}{107}},
  \bibinfo{pages}{015009} (\bibinfo{year}{2023}), \eprint{2209.06225}.

\bibitem[{\citenamefont{Hao et~al.}(2023)\citenamefont{Hao, Kansal, Duarte, and
  Chernyavskaya}}]{Hao:2022zns}
\bibinfo{author}{\bibfnamefont{Z.}~\bibnamefont{Hao}},
  \bibinfo{author}{\bibfnamefont{R.}~\bibnamefont{Kansal}},
  \bibinfo{author}{\bibfnamefont{J.}~\bibnamefont{Duarte}}, \bibnamefont{and}
  \bibinfo{author}{\bibfnamefont{N.}~\bibnamefont{Chernyavskaya}},
  \bibinfo{journal}{Eur. Phys. J. C} \textbf{\bibinfo{volume}{83}},
  \bibinfo{pages}{485} (\bibinfo{year}{2023}), \eprint{2212.07347}.

\bibitem[{\citenamefont{Golling et~al.}(2023)}]{Golling:2023juz}
\bibinfo{author}{\bibfnamefont{T.}~\bibnamefont{Golling}} \bibnamefont{et~al.},
  in \emph{\bibinfo{booktitle}{{34th Conference on Neural Information
  Processing Systems}}} (\bibinfo{year}{2023}), \eprint{2303.14134}.

\bibitem[{\citenamefont{Aad et~al.}(2023{\natexlab{a}})}]{ATLAS:2023azi}
\bibinfo{author}{\bibfnamefont{G.}~\bibnamefont{Aad}} \bibnamefont{et~al.}
  (\bibinfo{collaboration}{ATLAS}), \bibinfo{journal}{Phys. Rev. D}
  \textbf{\bibinfo{volume}{108}}, \bibinfo{pages}{052009}
  (\bibinfo{year}{2023}{\natexlab{a}}), \eprint{2306.03637}.

\bibitem[{\citenamefont{Metodiev et~al.}(2024)\citenamefont{Metodiev, Thaler,
  and Wynne}}]{Metodiev:2023izu}
\bibinfo{author}{\bibfnamefont{E.~M.} \bibnamefont{Metodiev}},
  \bibinfo{author}{\bibfnamefont{J.}~\bibnamefont{Thaler}}, \bibnamefont{and}
  \bibinfo{author}{\bibfnamefont{R.}~\bibnamefont{Wynne}},
  \bibinfo{journal}{Phys. Rev. D} \textbf{\bibinfo{volume}{110}},
  \bibinfo{pages}{055012} (\bibinfo{year}{2024}), \eprint{2312.00119}.

\bibitem[{\citenamefont{Sengupta et~al.}(2024)\citenamefont{Sengupta, Leigh,
  Raine, Klein, and Golling}}]{Sengupta:2023vtm}
\bibinfo{author}{\bibfnamefont{D.}~\bibnamefont{Sengupta}},
  \bibinfo{author}{\bibfnamefont{M.}~\bibnamefont{Leigh}},
  \bibinfo{author}{\bibfnamefont{J.~A.} \bibnamefont{Raine}},
  \bibinfo{author}{\bibfnamefont{S.}~\bibnamefont{Klein}}, \bibnamefont{and}
  \bibinfo{author}{\bibfnamefont{T.}~\bibnamefont{Golling}},
  \bibinfo{journal}{JHEP} \textbf{\bibinfo{volume}{04}}, \bibinfo{pages}{109}
  (\bibinfo{year}{2024}), \eprint{2312.10130}.

\bibitem[{\citenamefont{Cheng et~al.}(2024)\citenamefont{Cheng, Singh, and
  Nachman}}]{Cheng:2024yig}
\bibinfo{author}{\bibfnamefont{C.~L.} \bibnamefont{Cheng}},
  \bibinfo{author}{\bibfnamefont{G.}~\bibnamefont{Singh}}, \bibnamefont{and}
  \bibinfo{author}{\bibfnamefont{B.}~\bibnamefont{Nachman}}
  (\bibinfo{year}{2024}), \eprint{2405.08889}.

\bibitem[{\citenamefont{Grosso}(2024)}]{Grosso:2024nho}
\bibinfo{author}{\bibfnamefont{G.}~\bibnamefont{Grosso}},
  \bibinfo{journal}{JHEP} \textbf{\bibinfo{volume}{12}}, \bibinfo{pages}{093}
  (\bibinfo{year}{2024}), \eprint{2407.01249}.

\bibitem[{\citenamefont{Duffy et~al.}(2024)\citenamefont{Duffy, Hassanshah,
  Jastrzebski, and Malik}}]{Duffy:2024zog}
\bibinfo{author}{\bibfnamefont{C.}~\bibnamefont{Duffy}},
  \bibinfo{author}{\bibfnamefont{M.}~\bibnamefont{Hassanshah}},
  \bibinfo{author}{\bibfnamefont{M.}~\bibnamefont{Jastrzebski}},
  \bibnamefont{and} \bibinfo{author}{\bibfnamefont{S.}~\bibnamefont{Malik}}
  (\bibinfo{year}{2024}), \eprint{2407.07961}.

\bibitem[{\citenamefont{Das and Shih}(2024)}]{Das:2024fwo}
\bibinfo{author}{\bibfnamefont{R.}~\bibnamefont{Das}} \bibnamefont{and}
  \bibinfo{author}{\bibfnamefont{D.}~\bibnamefont{Shih}}
  (\bibinfo{year}{2024}), \eprint{2410.20537}.

\bibitem[{\citenamefont{Craig et~al.}(2024)\citenamefont{Craig, Howard, and
  Li}}]{Craig:2024rlv}
\bibinfo{author}{\bibfnamefont{N.}~\bibnamefont{Craig}},
  \bibinfo{author}{\bibfnamefont{J.~N.} \bibnamefont{Howard}},
  \bibnamefont{and} \bibinfo{author}{\bibfnamefont{H.}~\bibnamefont{Li}}
  (\bibinfo{year}{2024}), \eprint{2401.15542}.

\bibitem[{\citenamefont{Araz and Spannowsky}(2024)}]{Araz:2024lsl}
\bibinfo{author}{\bibfnamefont{J.~Y.} \bibnamefont{Araz}} \bibnamefont{and}
  \bibinfo{author}{\bibfnamefont{M.}~\bibnamefont{Spannowsky}}
  (\bibinfo{year}{2024}), \eprint{2409.04519}.

\bibitem[{\citenamefont{Das et~al.}(2024)\citenamefont{Das, Finke, Hein,
  Kasieczka, Kr\"amer, M\"uck, and Shih}}]{Das:2024eie}
\bibinfo{author}{\bibfnamefont{R.}~\bibnamefont{Das}},
  \bibinfo{author}{\bibfnamefont{T.}~\bibnamefont{Finke}},
  \bibinfo{author}{\bibfnamefont{M.}~\bibnamefont{Hein}},
  \bibinfo{author}{\bibfnamefont{G.}~\bibnamefont{Kasieczka}},
  \bibinfo{author}{\bibfnamefont{M.}~\bibnamefont{Kr\"amer}},
  \bibinfo{author}{\bibfnamefont{A.}~\bibnamefont{M\"uck}}, \bibnamefont{and}
  \bibinfo{author}{\bibfnamefont{D.}~\bibnamefont{Shih}}
  (\bibinfo{year}{2024}), \eprint{2411.00085}.

\bibitem[{\citenamefont{Hammad et~al.}(2024)\citenamefont{Hammad, Nojiri, and
  Yamazaki}}]{Hammad:2024dsn}
\bibinfo{author}{\bibfnamefont{A.}~\bibnamefont{Hammad}},
  \bibinfo{author}{\bibfnamefont{M.~M.} \bibnamefont{Nojiri}},
  \bibnamefont{and} \bibinfo{author}{\bibfnamefont{M.}~\bibnamefont{Yamazaki}}
  (\bibinfo{year}{2024}), \eprint{2411.09927}.

\bibitem[{\citenamefont{Farina et~al.}(2020)\citenamefont{Farina, Nakai, and
  Shih}}]{Farina:2018fyg}
\bibinfo{author}{\bibfnamefont{M.}~\bibnamefont{Farina}},
  \bibinfo{author}{\bibfnamefont{Y.}~\bibnamefont{Nakai}}, \bibnamefont{and}
  \bibinfo{author}{\bibfnamefont{D.}~\bibnamefont{Shih}},
  \bibinfo{journal}{Phys. Rev. D} \textbf{\bibinfo{volume}{101}},
  \bibinfo{pages}{075021} (\bibinfo{year}{2020}), \eprint{1808.08992}.

\bibitem[{\citenamefont{Heimel et~al.}(2019)\citenamefont{Heimel, Kasieczka,
  Plehn, and Thompson}}]{Heimel:2018mkt}
\bibinfo{author}{\bibfnamefont{T.}~\bibnamefont{Heimel}},
  \bibinfo{author}{\bibfnamefont{G.}~\bibnamefont{Kasieczka}},
  \bibinfo{author}{\bibfnamefont{T.}~\bibnamefont{Plehn}}, \bibnamefont{and}
  \bibinfo{author}{\bibfnamefont{J.~M.} \bibnamefont{Thompson}},
  \bibinfo{journal}{SciPost Phys.} \textbf{\bibinfo{volume}{6}},
  \bibinfo{pages}{030} (\bibinfo{year}{2019}), \eprint{1808.08979}.

\bibitem[{\citenamefont{Roy and Vijay}(2019)}]{Roy:2019jae}
\bibinfo{author}{\bibfnamefont{T.~S.} \bibnamefont{Roy}} \bibnamefont{and}
  \bibinfo{author}{\bibfnamefont{A.~H.} \bibnamefont{Vijay}}
  (\bibinfo{year}{2019}), \eprint{1903.02032}.

\bibitem[{\citenamefont{Cerri et~al.}(2019)\citenamefont{Cerri, Nguyen,
  Pierini, Spiropulu, and Vlimant}}]{Cerri:2018anq}
\bibinfo{author}{\bibfnamefont{O.}~\bibnamefont{Cerri}},
  \bibinfo{author}{\bibfnamefont{T.~Q.} \bibnamefont{Nguyen}},
  \bibinfo{author}{\bibfnamefont{M.}~\bibnamefont{Pierini}},
  \bibinfo{author}{\bibfnamefont{M.}~\bibnamefont{Spiropulu}},
  \bibnamefont{and} \bibinfo{author}{\bibfnamefont{J.-R.}
  \bibnamefont{Vlimant}}, \bibinfo{journal}{JHEP}
  \textbf{\bibinfo{volume}{05}}, \bibinfo{pages}{036} (\bibinfo{year}{2019}),
  \eprint{1811.10276}.

\bibitem[{\citenamefont{Blance et~al.}(2019)\citenamefont{Blance, Spannowsky,
  and Waite}}]{Blance:2019ibf}
\bibinfo{author}{\bibfnamefont{A.}~\bibnamefont{Blance}},
  \bibinfo{author}{\bibfnamefont{M.}~\bibnamefont{Spannowsky}},
  \bibnamefont{and} \bibinfo{author}{\bibfnamefont{P.}~\bibnamefont{Waite}},
  \bibinfo{journal}{JHEP} \textbf{\bibinfo{volume}{10}}, \bibinfo{pages}{047}
  (\bibinfo{year}{2019}), \eprint{1905.10384}.

\bibitem[{\citenamefont{van Beekveld et~al.}(2021)\citenamefont{van Beekveld,
  Caron, Hendriks, Jackson, Leinweber, Otten, Patrick, Ruiz De~Austri, Santoni,
  and White}}]{vanBeekveld:2020txa}
\bibinfo{author}{\bibfnamefont{M.}~\bibnamefont{van Beekveld}},
  \bibinfo{author}{\bibfnamefont{S.}~\bibnamefont{Caron}},
  \bibinfo{author}{\bibfnamefont{L.}~\bibnamefont{Hendriks}},
  \bibinfo{author}{\bibfnamefont{P.}~\bibnamefont{Jackson}},
  \bibinfo{author}{\bibfnamefont{A.}~\bibnamefont{Leinweber}},
  \bibinfo{author}{\bibfnamefont{S.}~\bibnamefont{Otten}},
  \bibinfo{author}{\bibfnamefont{R.}~\bibnamefont{Patrick}},
  \bibinfo{author}{\bibfnamefont{R.}~\bibnamefont{Ruiz De~Austri}},
  \bibinfo{author}{\bibfnamefont{M.}~\bibnamefont{Santoni}}, \bibnamefont{and}
  \bibinfo{author}{\bibfnamefont{M.}~\bibnamefont{White}},
  \bibinfo{journal}{JHEP} \textbf{\bibinfo{volume}{09}}, \bibinfo{pages}{024}
  (\bibinfo{year}{2021}), \eprint{2010.07940}.

\bibitem[{\citenamefont{Batson et~al.}(2021)\citenamefont{Batson, Haaf, Kahn,
  and Roberts}}]{Batson:2021agz}
\bibinfo{author}{\bibfnamefont{J.}~\bibnamefont{Batson}},
  \bibinfo{author}{\bibfnamefont{C.~G.} \bibnamefont{Haaf}},
  \bibinfo{author}{\bibfnamefont{Y.}~\bibnamefont{Kahn}}, \bibnamefont{and}
  \bibinfo{author}{\bibfnamefont{D.~A.} \bibnamefont{Roberts}},
  \bibinfo{journal}{JHEP} \textbf{\bibinfo{volume}{04}}, \bibinfo{pages}{280}
  (\bibinfo{year}{2021}), \eprint{2102.08380}.

\bibitem[{\citenamefont{Dillon et~al.}(2021)\citenamefont{Dillon, Plehn, Sauer,
  and Sorrenson}}]{Dillon:2021nxw}
\bibinfo{author}{\bibfnamefont{B.~M.} \bibnamefont{Dillon}},
  \bibinfo{author}{\bibfnamefont{T.}~\bibnamefont{Plehn}},
  \bibinfo{author}{\bibfnamefont{C.}~\bibnamefont{Sauer}}, \bibnamefont{and}
  \bibinfo{author}{\bibfnamefont{P.}~\bibnamefont{Sorrenson}},
  \bibinfo{journal}{SciPost Phys.} \textbf{\bibinfo{volume}{11}},
  \bibinfo{pages}{061} (\bibinfo{year}{2021}), \eprint{2104.08291}.

\bibitem[{\citenamefont{Finke et~al.}(2021)\citenamefont{Finke, Kr\"amer,
  Morandini, M\"uck, and Oleksiyuk}}]{Finke:2021sdf}
\bibinfo{author}{\bibfnamefont{T.}~\bibnamefont{Finke}},
  \bibinfo{author}{\bibfnamefont{M.}~\bibnamefont{Kr\"amer}},
  \bibinfo{author}{\bibfnamefont{A.}~\bibnamefont{Morandini}},
  \bibinfo{author}{\bibfnamefont{A.}~\bibnamefont{M\"uck}}, \bibnamefont{and}
  \bibinfo{author}{\bibfnamefont{I.}~\bibnamefont{Oleksiyuk}},
  \bibinfo{journal}{JHEP} \textbf{\bibinfo{volume}{06}}, \bibinfo{pages}{161}
  (\bibinfo{year}{2021}), \eprint{2104.09051}.

\bibitem[{\citenamefont{Atkinson et~al.}(2021)\citenamefont{Atkinson, Bhardwaj,
  Englert, Ngairangbam, and Spannowsky}}]{Atkinson:2021nlt}
\bibinfo{author}{\bibfnamefont{O.}~\bibnamefont{Atkinson}},
  \bibinfo{author}{\bibfnamefont{A.}~\bibnamefont{Bhardwaj}},
  \bibinfo{author}{\bibfnamefont{C.}~\bibnamefont{Englert}},
  \bibinfo{author}{\bibfnamefont{V.~S.} \bibnamefont{Ngairangbam}},
  \bibnamefont{and}
  \bibinfo{author}{\bibfnamefont{M.}~\bibnamefont{Spannowsky}},
  \bibinfo{journal}{JHEP} \textbf{\bibinfo{volume}{08}}, \bibinfo{pages}{080}
  (\bibinfo{year}{2021}), \eprint{2105.07988}.

\bibitem[{\citenamefont{Govorkova et~al.}(2022)}]{Govorkova:2021utb}
\bibinfo{author}{\bibfnamefont{E.}~\bibnamefont{Govorkova}}
  \bibnamefont{et~al.}, \bibinfo{journal}{Nature Mach. Intell.}
  \textbf{\bibinfo{volume}{4}}, \bibinfo{pages}{154} (\bibinfo{year}{2022}),
  \eprint{2108.03986}.

\bibitem[{\citenamefont{Fraser et~al.}(2022)\citenamefont{Fraser, Homiller,
  Mishra, Ostdiek, and Schwartz}}]{Fraser:2021lxm}
\bibinfo{author}{\bibfnamefont{K.}~\bibnamefont{Fraser}},
  \bibinfo{author}{\bibfnamefont{S.}~\bibnamefont{Homiller}},
  \bibinfo{author}{\bibfnamefont{R.~K.} \bibnamefont{Mishra}},
  \bibinfo{author}{\bibfnamefont{B.}~\bibnamefont{Ostdiek}}, \bibnamefont{and}
  \bibinfo{author}{\bibfnamefont{M.~D.} \bibnamefont{Schwartz}},
  \bibinfo{journal}{JHEP} \textbf{\bibinfo{volume}{03}}, \bibinfo{pages}{066}
  (\bibinfo{year}{2022}), \eprint{2110.06948}.

\bibitem[{\citenamefont{Tsan et~al.}(2021)\citenamefont{Tsan, Kansal, Aportela,
  Diaz, Duarte, Krishna, Mokhtar, Vlimant, and Pierini}}]{Tsan:2021brw}
\bibinfo{author}{\bibfnamefont{S.}~\bibnamefont{Tsan}},
  \bibinfo{author}{\bibfnamefont{R.}~\bibnamefont{Kansal}},
  \bibinfo{author}{\bibfnamefont{A.}~\bibnamefont{Aportela}},
  \bibinfo{author}{\bibfnamefont{D.}~\bibnamefont{Diaz}},
  \bibinfo{author}{\bibfnamefont{J.}~\bibnamefont{Duarte}},
  \bibinfo{author}{\bibfnamefont{S.}~\bibnamefont{Krishna}},
  \bibinfo{author}{\bibfnamefont{F.}~\bibnamefont{Mokhtar}},
  \bibinfo{author}{\bibfnamefont{J.-R.} \bibnamefont{Vlimant}},
  \bibnamefont{and} \bibinfo{author}{\bibfnamefont{M.}~\bibnamefont{Pierini}},
  in \emph{\bibinfo{booktitle}{{35th Conference on Neural Information
  Processing Systems}}} (\bibinfo{year}{2021}), \eprint{2111.12849}.

\bibitem[{\citenamefont{Jawahar et~al.}(2022)\citenamefont{Jawahar, Aarrestad,
  Chernyavskaya, Pierini, Wozniak, Ngadiuba, Duarte, and
  Tsan}}]{Jawahar:2021vyu}
\bibinfo{author}{\bibfnamefont{P.}~\bibnamefont{Jawahar}},
  \bibinfo{author}{\bibfnamefont{T.}~\bibnamefont{Aarrestad}},
  \bibinfo{author}{\bibfnamefont{N.}~\bibnamefont{Chernyavskaya}},
  \bibinfo{author}{\bibfnamefont{M.}~\bibnamefont{Pierini}},
  \bibinfo{author}{\bibfnamefont{K.~A.} \bibnamefont{Wozniak}},
  \bibinfo{author}{\bibfnamefont{J.}~\bibnamefont{Ngadiuba}},
  \bibinfo{author}{\bibfnamefont{J.}~\bibnamefont{Duarte}}, \bibnamefont{and}
  \bibinfo{author}{\bibfnamefont{S.}~\bibnamefont{Tsan}},
  \bibinfo{journal}{Front. Big Data} \textbf{\bibinfo{volume}{5}},
  \bibinfo{pages}{803685} (\bibinfo{year}{2022}), \eprint{2110.08508}.

\bibitem[{\citenamefont{Canelli et~al.}(2022)\citenamefont{Canelli, de~Cosa,
  Pottier, Niedziela, Pedro, and Pierini}}]{Canelli:2021aps}
\bibinfo{author}{\bibfnamefont{F.}~\bibnamefont{Canelli}},
  \bibinfo{author}{\bibfnamefont{A.}~\bibnamefont{de~Cosa}},
  \bibinfo{author}{\bibfnamefont{L.~L.} \bibnamefont{Pottier}},
  \bibinfo{author}{\bibfnamefont{J.}~\bibnamefont{Niedziela}},
  \bibinfo{author}{\bibfnamefont{K.}~\bibnamefont{Pedro}}, \bibnamefont{and}
  \bibinfo{author}{\bibfnamefont{M.}~\bibnamefont{Pierini}},
  \bibinfo{journal}{JHEP} \textbf{\bibinfo{volume}{02}}, \bibinfo{pages}{074}
  (\bibinfo{year}{2022}), \eprint{2112.02864}.

\bibitem[{\citenamefont{Ngairangbam et~al.}(2022)\citenamefont{Ngairangbam,
  Spannowsky, and Takeuchi}}]{Ngairangbam:2021yma}
\bibinfo{author}{\bibfnamefont{V.~S.} \bibnamefont{Ngairangbam}},
  \bibinfo{author}{\bibfnamefont{M.}~\bibnamefont{Spannowsky}},
  \bibnamefont{and} \bibinfo{author}{\bibfnamefont{M.}~\bibnamefont{Takeuchi}},
  \bibinfo{journal}{Phys. Rev. D} \textbf{\bibinfo{volume}{105}},
  \bibinfo{pages}{095004} (\bibinfo{year}{2022}), \eprint{2112.04958}.

\bibitem[{\citenamefont{Alvi et~al.}(2023)\citenamefont{Alvi, Bauer, and
  Nachman}}]{Alvi:2022fkk}
\bibinfo{author}{\bibfnamefont{S.}~\bibnamefont{Alvi}},
  \bibinfo{author}{\bibfnamefont{C.~W.} \bibnamefont{Bauer}}, \bibnamefont{and}
  \bibinfo{author}{\bibfnamefont{B.}~\bibnamefont{Nachman}},
  \bibinfo{journal}{JHEP} \textbf{\bibinfo{volume}{02}}, \bibinfo{pages}{220}
  (\bibinfo{year}{2023}), \eprint{2206.08391}.

\bibitem[{\citenamefont{Atkinson et~al.}(2022)\citenamefont{Atkinson, Bhardwaj,
  Englert, Konar, Ngairangbam, and Spannowsky}}]{Atkinson:2022uzb}
\bibinfo{author}{\bibfnamefont{O.}~\bibnamefont{Atkinson}},
  \bibinfo{author}{\bibfnamefont{A.}~\bibnamefont{Bhardwaj}},
  \bibinfo{author}{\bibfnamefont{C.}~\bibnamefont{Englert}},
  \bibinfo{author}{\bibfnamefont{P.}~\bibnamefont{Konar}},
  \bibinfo{author}{\bibfnamefont{V.~S.} \bibnamefont{Ngairangbam}},
  \bibnamefont{and}
  \bibinfo{author}{\bibfnamefont{M.}~\bibnamefont{Spannowsky}},
  \bibinfo{journal}{Front. Artif. Intell.} \textbf{\bibinfo{volume}{5}},
  \bibinfo{pages}{943135} (\bibinfo{year}{2022}), \eprint{2204.12231}.

\bibitem[{\citenamefont{Dillon et~al.}(2023)\citenamefont{Dillon, Favaro,
  Plehn, Sorrenson, and Kr\"amer}}]{Dillon:2022mkq}
\bibinfo{author}{\bibfnamefont{B.~M.} \bibnamefont{Dillon}},
  \bibinfo{author}{\bibfnamefont{L.}~\bibnamefont{Favaro}},
  \bibinfo{author}{\bibfnamefont{T.}~\bibnamefont{Plehn}},
  \bibinfo{author}{\bibfnamefont{P.}~\bibnamefont{Sorrenson}},
  \bibnamefont{and} \bibinfo{author}{\bibfnamefont{M.}~\bibnamefont{Kr\"amer}},
  \bibinfo{journal}{SciPost Phys. Core} \textbf{\bibinfo{volume}{6}},
  \bibinfo{pages}{074} (\bibinfo{year}{2023}), \eprint{2206.14225}.

\bibitem[{\citenamefont{Bhattacherjee et~al.}(2023)\citenamefont{Bhattacherjee,
  Konar, Ngairangbam, and Solanki}}]{Bhattacherjee:2023evs}
\bibinfo{author}{\bibfnamefont{B.}~\bibnamefont{Bhattacherjee}},
  \bibinfo{author}{\bibfnamefont{P.}~\bibnamefont{Konar}},
  \bibinfo{author}{\bibfnamefont{V.~S.} \bibnamefont{Ngairangbam}},
  \bibnamefont{and} \bibinfo{author}{\bibfnamefont{P.}~\bibnamefont{Solanki}}
  (\bibinfo{year}{2023}), \eprint{2308.13611}.

\bibitem[{\citenamefont{Aad et~al.}(2023{\natexlab{b}})}]{ATLAS:2022qxm}
\bibinfo{author}{\bibfnamefont{G.}~\bibnamefont{Aad}} \bibnamefont{et~al.}
  (\bibinfo{collaboration}{ATLAS}), \bibinfo{journal}{Eur. Phys. J. C}
  \textbf{\bibinfo{volume}{83}}, \bibinfo{pages}{681}
  (\bibinfo{year}{2023}{\natexlab{b}}), \eprint{2211.16345}.

\bibitem[{CMS(2016)}]{CMS-PAS-BTV-16-001}
\bibinfo{type}{Tech. Rep.}, \bibinfo{institution}{CERN},
  \bibinfo{address}{Geneva} (\bibinfo{year}{2016}),
  \urlprefix\url{https://cds.cern.ch/record/2205149}.

\bibitem[{\citenamefont{Munkres}(2000)}]{Munkres2000Topology}
\bibinfo{author}{\bibfnamefont{J.~R.} \bibnamefont{Munkres}},
  \emph{\bibinfo{title}{{Topology}}} (\bibinfo{publisher}{Prentice Hall, Inc.},
  \bibinfo{year}{2000}), \bibinfo{edition}{2nd} ed., ISBN
  \bibinfo{isbn}{0131816292},
  \urlprefix\url{http://www.worldcat.org/isbn/0131816292}.

\bibitem[{\citenamefont{Whitney}(1936)}]{c992e7fe-9a8a-3576-849e-c8a0d8dab89c}
\bibinfo{author}{\bibfnamefont{H.}~\bibnamefont{Whitney}},
  \bibinfo{journal}{Annals of Mathematics} \textbf{\bibinfo{volume}{37}},
  \bibinfo{pages}{645} (\bibinfo{year}{1936}), ISSN \bibinfo{issn}{0003486X,
  19398980}, \urlprefix\url{http://www.jstor.org/stable/1968482}.

\bibitem[{\citenamefont{Whitney}(1944)}]{38b777be-fb0a-3998-b108-baf022f5dfd4}
\bibinfo{author}{\bibfnamefont{H.}~\bibnamefont{Whitney}},
  \bibinfo{journal}{Annals of Mathematics} \textbf{\bibinfo{volume}{45}},
  \bibinfo{pages}{220} (\bibinfo{year}{1944}), ISSN \bibinfo{issn}{0003486X,
  19398980}, \urlprefix\url{http://www.jstor.org/stable/1969265}.

\bibitem[{\citenamefont{Alwall et~al.}(2014)\citenamefont{Alwall, Frederix,
  Frixione, Hirschi, Maltoni, Mattelaer, Shao, Stelzer, Torrielli, and
  Zaro}}]{Alwall:2014hca}
\bibinfo{author}{\bibfnamefont{J.}~\bibnamefont{Alwall}},
  \bibinfo{author}{\bibfnamefont{R.}~\bibnamefont{Frederix}},
  \bibinfo{author}{\bibfnamefont{S.}~\bibnamefont{Frixione}},
  \bibinfo{author}{\bibfnamefont{V.}~\bibnamefont{Hirschi}},
  \bibinfo{author}{\bibfnamefont{F.}~\bibnamefont{Maltoni}},
  \bibinfo{author}{\bibfnamefont{O.}~\bibnamefont{Mattelaer}},
  \bibinfo{author}{\bibfnamefont{H.~S.} \bibnamefont{Shao}},
  \bibinfo{author}{\bibfnamefont{T.}~\bibnamefont{Stelzer}},
  \bibinfo{author}{\bibfnamefont{P.}~\bibnamefont{Torrielli}},
  \bibnamefont{and} \bibinfo{author}{\bibfnamefont{M.}~\bibnamefont{Zaro}},
  \bibinfo{journal}{JHEP} \textbf{\bibinfo{volume}{07}}, \bibinfo{pages}{079}
  (\bibinfo{year}{2014}), \eprint{1405.0301}.

\bibitem[{\citenamefont{Bierlich et~al.}(2022)}]{Bierlich:2022pfr}
\bibinfo{author}{\bibfnamefont{C.}~\bibnamefont{Bierlich}}
  \bibnamefont{et~al.}, \bibinfo{journal}{SciPost Phys. Codeb.}
  \textbf{\bibinfo{volume}{2022}}, \bibinfo{pages}{8} (\bibinfo{year}{2022}),
  \eprint{2203.11601}.

\bibitem[{\citenamefont{Brivio et~al.}(2017)\citenamefont{Brivio, Jiang, and
  Trott}}]{Brivio:2017btx}
\bibinfo{author}{\bibfnamefont{I.}~\bibnamefont{Brivio}},
  \bibinfo{author}{\bibfnamefont{Y.}~\bibnamefont{Jiang}}, \bibnamefont{and}
  \bibinfo{author}{\bibfnamefont{M.}~\bibnamefont{Trott}},
  \bibinfo{journal}{JHEP} \textbf{\bibinfo{volume}{12}}, \bibinfo{pages}{070}
  (\bibinfo{year}{2017}), \eprint{1709.06492}.

\bibitem[{\citenamefont{Brivio}(2021)}]{Brivio:2020onw}
\bibinfo{author}{\bibfnamefont{I.}~\bibnamefont{Brivio}},
  \bibinfo{journal}{JHEP} \textbf{\bibinfo{volume}{04}}, \bibinfo{pages}{073}
  (\bibinfo{year}{2021}), \eprint{2012.11343}.

\bibitem[{\citenamefont{Cacciari et~al.}(2008)\citenamefont{Cacciari, Salam,
  and Soyez}}]{Cacciari:2008gp}
\bibinfo{author}{\bibfnamefont{M.}~\bibnamefont{Cacciari}},
  \bibinfo{author}{\bibfnamefont{G.~P.} \bibnamefont{Salam}}, \bibnamefont{and}
  \bibinfo{author}{\bibfnamefont{G.}~\bibnamefont{Soyez}},
  \bibinfo{journal}{JHEP} \textbf{\bibinfo{volume}{04}}, \bibinfo{pages}{063}
  (\bibinfo{year}{2008}), \eprint{0802.1189}.

\bibitem[{\citenamefont{Cacciari et~al.}(2012)\citenamefont{Cacciari, Salam,
  and Soyez}}]{Cacciari:2011ma}
\bibinfo{author}{\bibfnamefont{M.}~\bibnamefont{Cacciari}},
  \bibinfo{author}{\bibfnamefont{G.~P.} \bibnamefont{Salam}}, \bibnamefont{and}
  \bibinfo{author}{\bibfnamefont{G.}~\bibnamefont{Soyez}},
  \bibinfo{journal}{Eur. Phys. J. C} \textbf{\bibinfo{volume}{72}},
  \bibinfo{pages}{1896} (\bibinfo{year}{2012}), \eprint{1111.6097}.

\bibitem[{\citenamefont{Pedregosa et~al.}(2011)\citenamefont{Pedregosa,
  Varoquaux, Gramfort, Michel, Thirion, Grisel, Blondel, Prettenhofer, Weiss,
  Dubourg et~al.}}]{scikit-learn}
\bibinfo{author}{\bibfnamefont{F.}~\bibnamefont{Pedregosa}},
  \bibinfo{author}{\bibfnamefont{G.}~\bibnamefont{Varoquaux}},
  \bibinfo{author}{\bibfnamefont{A.}~\bibnamefont{Gramfort}},
  \bibinfo{author}{\bibfnamefont{V.}~\bibnamefont{Michel}},
  \bibinfo{author}{\bibfnamefont{B.}~\bibnamefont{Thirion}},
  \bibinfo{author}{\bibfnamefont{O.}~\bibnamefont{Grisel}},
  \bibinfo{author}{\bibfnamefont{M.}~\bibnamefont{Blondel}},
  \bibinfo{author}{\bibfnamefont{P.}~\bibnamefont{Prettenhofer}},
  \bibinfo{author}{\bibfnamefont{R.}~\bibnamefont{Weiss}},
  \bibinfo{author}{\bibfnamefont{V.}~\bibnamefont{Dubourg}},
  \bibnamefont{et~al.}, \bibinfo{journal}{Journal of Machine Learning Research}
  \textbf{\bibinfo{volume}{12}}, \bibinfo{pages}{2825} (\bibinfo{year}{2011}).

\bibitem[{\citenamefont{Paszke et~al.}(2019)\citenamefont{Paszke, Gross, Massa,
  Lerer, Bradbury, Chanan, Killeen, Lin, Gimelshein, Antiga
  et~al.}}]{10.5555/3454287.3455008}
\bibinfo{author}{\bibfnamefont{A.}~\bibnamefont{Paszke}},
  \bibinfo{author}{\bibfnamefont{S.}~\bibnamefont{Gross}},
  \bibinfo{author}{\bibfnamefont{F.}~\bibnamefont{Massa}},
  \bibinfo{author}{\bibfnamefont{A.}~\bibnamefont{Lerer}},
  \bibinfo{author}{\bibfnamefont{J.}~\bibnamefont{Bradbury}},
  \bibinfo{author}{\bibfnamefont{G.}~\bibnamefont{Chanan}},
  \bibinfo{author}{\bibfnamefont{T.}~\bibnamefont{Killeen}},
  \bibinfo{author}{\bibfnamefont{Z.}~\bibnamefont{Lin}},
  \bibinfo{author}{\bibfnamefont{N.}~\bibnamefont{Gimelshein}},
  \bibinfo{author}{\bibfnamefont{L.}~\bibnamefont{Antiga}},
  \bibnamefont{et~al.}, \emph{\bibinfo{title}{PyTorch: an imperative style,
  high-performance deep learning library}} (\bibinfo{publisher}{Curran
  Associates Inc.}, \bibinfo{address}{Red Hook, NY, USA},
  \bibinfo{year}{2019}).

\bibitem[{\citenamefont{Kingma and Ba}(2015)}]{DBLP:journals/corr/KingmaB14}
\bibinfo{author}{\bibfnamefont{D.~P.} \bibnamefont{Kingma}} \bibnamefont{and}
  \bibinfo{author}{\bibfnamefont{J.}~\bibnamefont{Ba}}, in
  \emph{\bibinfo{booktitle}{3rd International Conference on Learning
  Representations, {ICLR} 2015, San Diego, CA, USA, May 7-9, 2015, Conference
  Track Proceedings}}, edited by
  \bibinfo{editor}{\bibfnamefont{Y.}~\bibnamefont{Bengio}} \bibnamefont{and}
  \bibinfo{editor}{\bibfnamefont{Y.}~\bibnamefont{LeCun}}
  (\bibinfo{year}{2015}), \urlprefix\url{http://arxiv.org/abs/1412.6980}.

\end{thebibliography}
%%%%%%%%%%%%%%%%%%%%%%%%%%%%%%%%%%%%%%%%%%%%%%%%%%

\end{document}